\documentclass[prl,twocolumn,noshowpacs,preprintnumbers,amsmath,amssymb,nofootinbib]{revtex4-2}

\usepackage{mathrsfs}
\usepackage{graphicx}
\usepackage{bm}
\usepackage{ascmac}
\usepackage[hidelinks]{hyperref}

\newcommand{\mcL}{\mathcal{L}}

\newcommand{\ra}{\rangle}
\newcommand{\la}{\langle}
\newcommand{\tb}{\textbf}

\newcommand{\eps}{\epsilon}

\newcommand{\dsgm}{\dot{\sigma}}

\newcommand{\xiCP}{\xi^{\mathrm{CP}}}

\newcommand{\rini}{\mathrm{ini}}
\newcommand{\mrcan}{\mathrm{can}}
\newcommand{\pd}{\partial}

\newcommand*\dd{\mathop{}\!\mathrm{d}}

\newcommand*\dt{\mathop{}\!\mathrm{d}t}

\newcommand*\ds{\mathop{}\!\mathrm{d}s}

\begin{document}
\title{	
		Stochastic thermodynamics for classical non-Markov jump processes
	   }
\author{Kiyoshi Kanazawa}
\author{Andreas Dechant}
\affiliation{
			Department of Physics, Graduate School of Science, Kyoto University, Kyoto 606-8502, Japan
		}
\date{August 4, 2026}

\begin{abstract}
	Stochastic thermodynamics investigates energetic and entropic bounds in small systems, but its foundational results rely on the Markov (memoryless) assumption. Although the Markov assumption is questionable in real experimental setups, extending stochastic thermodynamics to general non-Markov systems has proven challenging. Fundamentally, it has been elusive how to model memory-dependent non-Gaussian fluctuations consistently with thermodynamic laws. Here we establish stochastic thermodynamics for classical non-Markov jump processes with equilibrium bath degrees of freedom. Building on Markov embedding, we introduce a key technique, called the {\it Fourier embedding}, which converts non-Markov jump processes into Markovian field dynamics of auxiliary Fourier modes. This yields necessary and sufficient conditions for time-reversal symmetry and enables the derivation of the second law for a broad class of Fourier-embeddable dynamics with strong memory. Remarkably, our cumulative entropy production is determined by observations of the target system alone, independently of the embedding. We illustrate our framework with two non-Markov models: (i)~a history-dependent two-state model and (ii)~a history-dependent random walk. Interestingly, these models can be interpreted as jump-process counterparts of Zwanzig's model and thus admit a transparent microscopic interpretation. Our work offers a guiding principle for thermodynamically consistent, physics-informed modelling of history-dependent fluctuations.
\end{abstract}

\maketitle
  Stochastic thermodynamics~\cite{ShiraishiB} provides a fundamental framework for characterising the energetics of small and noisy systems across diverse areas \cite{Ciliberto2017} (e.g., biophysics, quantum nanotechnology, and thermodynamic computing). Despite its successes, its applicability is significantly constrained by its core assumption of memoryless Markovian fluctuations, since the theory traditionally relies on the mathematics of Markov processes, in which fluctuations do not depend on the system's history~\cite{GardinerB}. In this work, we lift this constraint and extend the framework to non-Markov systems with strong memory. 
    
  It has been well-argued that non-Markovianity is ubiquitous in various systems (e.g., physical~\cite{KuboB,ZwanzigB}, biological~\cite{KanazawaNature2020}, chemical~\cite{VanKampen}, neural~\cite{Gerhard2017}, and social systems~\cite{Hawkes,Crane2008}), where fluctuations depend explicitly on the system's entire history. Many physicists have recognised this conceptual problem for a long time: the Markov assumption is incorrect even for the prototypical Brownian motion in water due to hydrodynamic memory~\cite{RaizenNPhys2011,RaizenScience2014}. However, precise modelling of such non-Markov processes is theoretically challenging, except for a few special classes (such as the generalised Langevin equation (GLE)~\cite{SpeckSeifert2007,Ohkuma2007,Munakata2014,Debiossac2022}, semi-Markov processes~\cite{KlafterB,MaesSemiMarkov2009,JannPRX2022}, weak-memory limits~\cite{Brandner2025,Brandner2025B}, and the generalised Glauber dynamics~\cite{ChenPRX2023}), and a stochastic-thermodynamic formulation has been missing for general non-Markov processes. Moreover, the presence of memory also causes fundamental issues, as time-reversal can lead to acausal features, where a time-reversed process depends on its own future~\cite{Munakata2014,Debiossac2022}. Thus, it remains unclear how to model non-Markovian fluctuations consistently with thermodynamic laws, potentially prohibiting further experimental development.

	Here we establish stochastic thermodynamics for non-Markov jump processes---jump processes with path-dependent intensities, going beyond the Markov or semi-Markov assumptions. We formulate the first and second laws for such genuinely strong-memory processes. To this end, we develop a new technique, called the {\it Fourier embedding}, to construct a master equation for the system coupled to an auxiliary field of Fourier modes. Since the resulting embedded dynamics is Markovian, we can specify its time-reversal symmetry and formulate stochastic work, heat, and entropy even in the presence of non-Markov jump noise. As a physically intuitive example, we illustrate our framework with the class of Zwanzig jump models: thermodynamically consistent, physically interpretable jump-process counterparts of Zwanzig's classical oscillator bath. This work opens up a new thermodynamic paradigm for small systems, offering key insights for the experimental modelling of thermal fluctuations with strong memory.

\begin{figure}
		\centering
		\includegraphics[width=87mm]{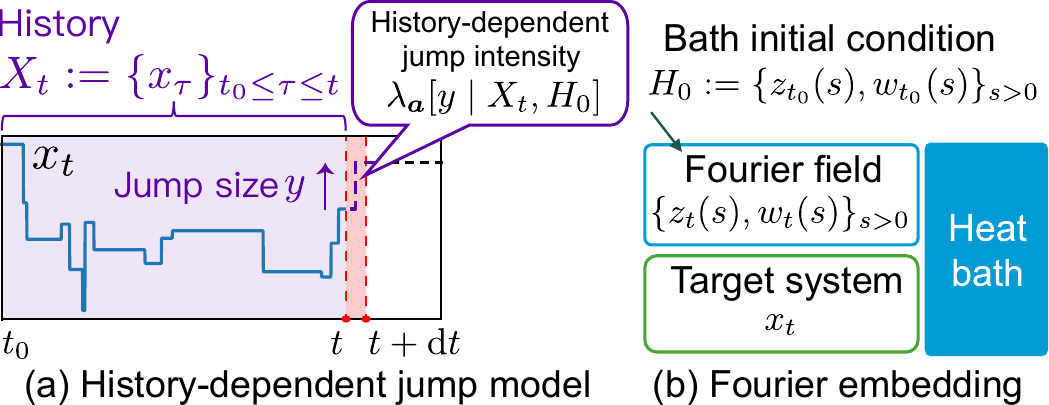}
		\caption{
			(a)~Non-Markov jump model with intensity $\lambda_{\bm a}[y\mid X_t, H_0]$ depending on the system's entire history $X_t:=\{x_{\tau}\}_{t_0\leq \tau\leq t}$ and the bath initial condition $H_0$.  (b)~Schematic of the Fourier embedding. In the extended phase space, the point $\Gamma_t:=(x_t,\{z_t(s),w_t(s)\}_s)$ follows the Markovian dynamics~\eqref{eq:set_complete_dynamics_hdCP} consistent with thermodynamic laws. By tracing out the auxiliary Fourier field $\{z_t(s),w_t(s)\}_s$, the target system $x_t$ follows the original non-Markov dynamics~\eqref{eq:def_history_dependent_Poisson}. The bath initial condition $H_0:=\{z_{t_0}(s),w_{t_0}(s)\}_{s>0}$ is sampled from the conditional canonical distribution given $x_{t_0}$.
		}
		\label{fig:schematic}
	\end{figure}

\paragraph{Model.}
	In this work, we study a one-dimensional history-dependent non-Markov jump process $x_t$ whose intensity depends on the system's history $X_t$ [see Fig.~\ref{fig:schematic}(a)]:
    \begin{equation}\label{eq:def_history_dependent_Poisson}
		\frac{\dd x_t}{\dt} = \xiCP_{\lambda_{\bm{a}}[y \mid X_t, H_0]}, \>\>\>
		X_t := \{x_{\tau}\}_{t_0\leq \tau\leq t},
	\end{equation}
	where $\bm{a}$ is the external control parameter and $\xiCP_{\lambda_{\bm{a}}[y \mid X_t, H_0]}$ is the compound Poisson noise. We also allow the intensity to depend on additional random variables $H_0$, defined in Eq.~\eqref{eq:def:InitCondition-H0}. As we discuss below, these are required for thermodynamic consistency and can be interpreted as the initial condition of the bath. For a jump size $y$, the intensity $\lambda_{\bm{a}}[y\mid X_t, H_0]$ depends on the system's full history $X_t$: the probability of observing a jump size in $[y,y+\dd y)$ during the time interval $[t,t+\dd t)$ is $\lambda_{\bm{a}}[y \mid X_t, H_0] \dd y \dd t$. We use square brackets for functionals to indicate that their arguments are function-valued; for example, $f(\{x_{\tau}\}_{t_0\leq \tau\leq t})=f[x]$.

	This model flexibly accommodates broad setups (e.g., a non-Markov two-state system; inspired by the semiclassical dynamics of quantum dots~\cite{Shimizu2001,Margolin2006}), but has not been studied in stochastic thermodynamics due to its mathematical difficulty. Several fundamental questions naturally arise: How can we set the intensity consistently with thermodynamics? Can we formulate the first and second laws? Can we construct thermodynamically consistent non-Markov models with a microscopic interpretation? In this work, we answer all these questions. 
	
\paragraph{Fourier embedding.}
	To solve this problem, we build on the established idea of Markov embedding and introduce a new technique, the {\it Fourier embedding}, for non-Markov stochastic thermodynamics. Markov embedding converts a low-dimensional non-Markov process into a high-dimensional Markov process by adding an appropriate set of auxiliary variables. This idea has a long history in classical statistical physics~\cite{ZwanzigB,Mori1965,Mori1965CF,Zwanzig1973,ErmakBuckholz1980,Klippenstein2021,Siegle2010,KzDidier2019PRL,KzDidier2021PRL,KzDidier2024PRR}, and closely related ideas were later developed in quantum open systems~\cite{Imamoglu1994,Garraway1997,Tamascelli2018,Lambert2019,Cirio2023,Menczel2024,Strasberg2016} (see SI for a brief review). By tracing out the auxiliary variables, the extended Markov dynamics reduces back to the original non-Markov one. This is physically intuitive---any stochastic model in physics is rooted in the coarse-graining of high-dimensional, Markovian microscopic dynamics [Fig.~\ref{fig:schematic}(b)]---so Markov embedding acts as an inverse of coarse-graining, providing an effective ``microscopic'' (Markovian) representation of a given non-Markovian model.

	Let us represent the pair $(X_t,H_0)$ using the Fourier transform of the velocity $v_{\tau} = \dd x_{\tau}/\dd \tau$ with an arbitrary initial time $t_0$: 
	\begin{subequations}			
		\label{def:Fourier_MarkovEmbedding_HDP}
		\begin{equation}
		\begin{aligned}
			z_t(s) :=& z_{t}^{\rini}(s) + \int_{t_0}^{t} \dd \tau v_{\tau}\sin \{s(t-\tau)\}, \\ 
			w_t(s) :=& w_{t}^{\rini}(s) + \int_{t_0}^{t} \dd \tau v_{\tau}\cos \{s(t-\tau)\},
		\end{aligned}
		\end{equation}
		where $z_{t}^{\rini}(s)$ and $w_{t}^{\rini}(s)$ are the contributions generated by the bath initial condition $H_0$. We then adopt the following variable set
		\begin{equation}
			\Gamma_t := (x_t, \{z_t(s), w_t(s)\}_{s>0}).
		\end{equation}
	\end{subequations}
	The time-reversals are defined as $x_t^* = \eps x_t$, $z_t^*(s)=-\eps z_t(s)$, and $w_t^*(s)=\eps w_t(s)$, where $(\cdot)^*$ denotes time reversal, $\eps=\pm1$ is the parity of $x$, and $\bm{a}^*$ is the time-reversed control parameter.

	This Fourier embedding converts the original non-Markov dynamics into the Markov dynamics of the auxiliary Fourier field variables [Fig.~\ref{fig:schematic}(b); see Methods]:
	\begin{subequations}			
		\label{eq:set_complete_dynamics_hdCP}
		\begin{equation}
		\begin{gathered}
			\frac{\dd x_t}{\dt} = \xiCP_{\lambda_{\bm{a}}[y\mid \Gamma_t]}, \>\>\> 
			\frac{\partial z_t(s)}{\partial t} = s w_t(s), \\
			\frac{\partial w_t(s)}{\partial t} = -s z_t(s) + \xiCP_{\lambda_{\bm{a}}[y\mid \Gamma_t]},
		\end{gathered}
		\end{equation}
        The jump term $\xiCP_{\lambda_{\bm{a}}[y \mid \Gamma_t]}$ acts on $x_t$ and $w_t(s)$ for all $s>0$ simultaneously. The contribution generated by the bath initial condition $H_0$ is given by
		\begin{equation}
			\label{eq:def:InitCondition-H0}
		\begin{gathered}
			H_0:=\{z_{t_0}(s),w_{t_0}(s)\}_{s>0}, \\
			z_{t}^{\rini}(s) \!=\! z_{t_0}(s)\!\cos \{s(t\!-\!t_0)\} \!+\! w_{t_0}(s)\!\sin \{s(t\!-\!t_0)\}, \\
			w_{t}^{\rini}(s) \!=\! w_{t_0}(s)\!\cos \{s(t\!-\!t_0)\} \!-\! z_{t_0}(s)\!\sin \{s(t\!-\!t_0)\}.
		\end{gathered}
		\end{equation}
	\end{subequations}
	The initial state $(x_{t_0},H_0)$ is sampled from the canonical distribution~\eqref{eq:Assumption1-can_dist} for thermodynamic consistency.

	In this work, a history-dependent jump process is called {\it Fourier-embeddable} if its intensity depends on $(X_t,H_0)$ only through $\Gamma_t$: $\lambda_{\bm{a}}[y\mid X_t,H_0]=\lambda_{\bm{a}}[y\mid \Gamma_t]$. We focus on this class of non-Markov jump processes. Also, $x_t$ and $\{z_t(s),w_t(s)\}_s$ are called the {\it target system} and {\it auxiliary Fourier field}, respectively. The velocity history is encoded in the Fourier field as an infinite collection of ``harmonic-oscillator'' bath modes. In quantum thermodynamics, related techniques are the pseudomode method~\cite{Imamoglu1994,Garraway1997} and reaction-coordinate mapping~\cite{Strasberg2016} for quantum systems linearly coupled to Gaussian baths. By contrast, our construction is formulated directly for classical history-dependent jump processes with nonlinear intensity functionals and also accommodates non-Gaussian auxiliary-field statistics.

\paragraph{Field master equation.}
	After the Fourier embedding, the original non-Markov dynamics is equivalent to the extended Markov dynamics. Therefore, we can derive the time evolution of the probability density functional (PDF) $P_t[\Gamma]$, which is described by the field master equation (fME; see Methods for its derivation):
	\begin{subequations}\label{eq:field_master_hdCP}
	\begin{equation}
		\frac{\partial P_t[\Gamma]}{\partial t} = \left(\mcL_{\rm A}+\mcL_{\rm J}\right)P_t[\Gamma].
	\end{equation}
	The advective and jump Liouville operators are given by 
	\begin{equation}
	\begin{aligned}
		&\mcL_{\rm A}P_t := 
		\int_0^\infty s\left\{z(s)\frac{\delta P_t[\Gamma]}{\delta w(s)} -w(s)\frac{\delta P_t[\Gamma]}{\delta z(s)}\right\}\ds , \\
		&\mcL_{\rm J}P_t := \!\! \int_{-\infty}^{\infty}\!\!\!\! \left\{\lambda_{\bm{a}}[y \>|\> \Gamma']P_t[\Gamma']\!-\!\lambda_{\bm{a}}[y \>|\> \Gamma]P_t[\Gamma]\right\}\dd y
	\end{aligned}
	\end{equation}
	\end{subequations}
	with $\Gamma':=\Gamma-\Delta \Gamma_y$ and $\Delta \Gamma_y:=(y,\{0\}_s,\{y\}_s)$.

    In the following, we assume that the fME~\eqref{eq:field_master_hdCP} admits a stationary PDF and that the initial condition $(x_{t_0},H_0)$ is sampled from the stationary PDF. We refer to this assumption as {\it Assumption 0}. As shown later, we present two non-Markov jump models that admit a stationary PDF and cannot be captured by the GLE and semi-Markov frameworks.

\paragraph{Time-reversal symmetry.}
	Now that the fME~\eqref{eq:field_master_hdCP} is derived, we can derive the necessary and sufficient conditions for its time-reversal symmetry. We make two key assumptions: For fixed $\bm{a}$, we assume the canonical equilibrium PDF with the total energy $E_{\bm{a}}[\Gamma]$ and the free energy $F(\bm{a})$ as the stationary solution of the fME~\eqref{eq:field_master_hdCP}: 
		\begin{subequations}\label{eq:Assumption1}
		\begin{align}\label{eq:Assumption1-can_dist}			
			P_{\mrcan;\bm{a}}[\Gamma] = e^{\beta (F(\bm{a})-E_{\bm{a}}[\Gamma])}
		\end{align}
		with the even-parity Hamiltonian $E_{\bm{a}}[\Gamma] = E_{\bm{a}^*}[\Gamma^*]$ and the inverse temperature $\beta$. Moreover, we assume detailed balance: 
		\begin{equation}\label{eq:Assumption1-DB}
			\frac{1}{\beta}\ln \frac{\lambda_{\bm{a}^*}\left[\eps(x-x')\mid\Gamma'^{*}\right]}{\lambda_{\bm{a}}\left[x'-x \mid \Gamma\right]} = E_{\bm{a}}[\Gamma']-E_{\bm{a}}[\Gamma].
		\end{equation}
	\end{subequations}
	These conditions~\eqref{eq:Assumption1} are called {\it Assumption 1} in this work. Assumption 1 is a natural generalisation of the time-reversal symmetry for Markov jump processes~\cite{ShiraishiB,GardinerB}; it provides an energetic interpretation of the dynamics of the phase-space point $\Gamma$. Also, we assume the time-reversal invariance of the total intensity as 	
	\begin{equation}\label{eq:Assumption2}
		\lambda_{{\rm tot};\bm{a}}[\Gamma] = \lambda_{{\rm tot};\bm{a}^*}[\Gamma^*], \>\>\> 
		\lambda_{{\rm tot};\bm{a}}[\Gamma] := \int_{-\infty}^\infty \!\!\lambda_{\bm{a}}[y \mid \Gamma]\dd y,
	\end{equation}
	which we call {\it Assumption 2}; this is a standard assumption customarily imposed even in Markov stochastic thermodynamics to ensure time-reversal consistency, especially in the presence of odd variables~\cite{ShiraishiB}. 
    
    The fME~\eqref{eq:field_master_hdCP} and Assumptions 1 and 2 constitute our first main result and provide the mathematical starting point for developing non-Markov stochastic thermodynamics: Assumptions 1 and 2 serve as a guiding principle for modelling memory-dependent non-Gaussian jumps consistently with thermodynamic laws. Remarkably, under Assumptions 1 and 2, the fME~\eqref{eq:field_master_hdCP} can be solved by the method of characteristics to obtain its general solution, which restricts the form of the energy functional (see Methods).  

\paragraph{Thermodynamic laws.}
	We formulate stochastic thermodynamics for the model~\eqref{eq:def_history_dependent_Poisson} under Assumptions 1 and 2. The first law is given by 
	\begin{subequations}
		\label{eq:first-law}
		\begin{equation}
			\dd E_{\bm{a}}[\Gamma] = \dd W + \dd Q,
		\end{equation}
		where the increments of stochastic work and heat are defined for the transition $\bm{a} \to \bm{a}+\dd \bm{a}$ and $\Gamma \to \Gamma'$ as
		\begin{align}
			\dd W \!:=\! \dd \bm{a}\frac{\pd E_{\bm{a}}[\Gamma]}{\pd \bm{a}}, \>\>
			\dd Q \!:=\! \frac{1}{\beta}\ln \frac{\lambda_{\bm{a}^*}[\eps(x-x')\!\mid\! \Gamma'^*]}{\lambda_{\bm{a}}[x'-x \!\mid\! \Gamma]}.
		\end{align}
	\end{subequations}
	Moreover, the second law for the total entropy production rate (EPR) $\dsgm_{\rm tot}$ reads 
	\begin{subequations}
		\label{eq:second-law}
		\begin{equation}
			\dsgm_{\rm tot} := \dsgm_{\rm sys} + \dsgm_{\rm bath} \geq 0.
		\end{equation}
		Here, the EPRs for the system and bath are defined as
		\begin{align}
			\dsgm_{\rm sys} := -\frac{\dd}{\dd t}\!\int\! \dd \Gamma P_t[\Gamma]\ln P_t[\Gamma] , \>\> \dsgm_{\rm bath} := - \beta \frac{\la \dd Q \ra}{\dd t},
		\end{align}
	\end{subequations}
    where $\dd \Gamma:=\dd x\prod_{s}\{\dd z(s)\dd w(s)\}$ denotes the volume element and $\la \dots\ra$ denotes the ensemble average.
    
	The first law~\eqref{eq:first-law} and the second law~\eqref{eq:second-law} are our second main result, constituting a pillar of non-Markov stochastic thermodynamics (see Methods for their derivation). We stress that Assumptions 0--2 are sufficient to model non-Markovian fluctuations consistently with thermodynamics, offering a guiding principle for model building. Also, the strict non-negativity of our EPR after Markov embedding is in sharp contrast to the conventional non-Markov EPR, which can be negative~\cite{Strasberg2019}.

\paragraph{Gauge invariance of entropy production.}
	Generally, Markov embedding is not unique: there can be several or even infinitely many different Markov embeddings that lead to the same non-Markov dynamics when focusing on the target system [see Refs.~\cite{KzDidier2019PRL,KzDidier2021PRL,KzDidier2024PRR} for the Laplace embedding as another candidate]. It is a natural question whether the second law is ``gauge invariant'', that is, whether the cumulative entropy production is independent of the arbitrary choice of Markov embedding. 
	
	Our answer relies on a structural condition on the energy:
	\begin{equation}
			E_{\bm{a}}[\Gamma]:= E_{{\rm ctrl};\bm{a}}(x) +  E_{\rm nctrl}\left(x,\{z(s),w(s)\}_s\right) \label{eq:controllable-energy},
	\end{equation}
	where $E_{{\rm ctrl};\bm{a}}(x)$ depends only on $x$ and $E_{\rm nctrl}$ is independent of $\bm{a}$. We refer to this condition as {\it Assumption 3}. Under Assumption 3, $\sigma_{\rm tot}$ admits a representation involving only the statistics of the target system; it is therefore (i)~accessible from observations of $x$ alone and (ii)~independent of the choice of embedding. 

	Let $P_{{\rm can};\bm{a}}(x)$ be the equilibrium marginal PDF of the target system at fixed $\bm{a}$, which involves no auxiliary variables. We define $E_{{\rm exp};\bm{a}}(x)$ and $F_{\rm exp}(\bm{a})$ through $P_{{\rm can};\bm{a}}(x):=e^{\beta[F_{\rm exp}(\bm{a})-E_{{\rm exp};\bm{a}}(x)]}$; both are thus fixed by measurements on $x$ alone. For any equilibrium-to-equilibrium transition, we show 
	\begin{equation}
		\label{eq:second-law-work}
			\sigma_{\rm tot}=\sigma_{\rm exp}:=\beta\left(\la\Delta W_{\rm exp}\ra-\Delta F_{\rm exp}\right)\geq 0
	\end{equation}
	with $\Delta W_{\rm exp}:=\int_{\bm{a}_{\rm ini}}^{\bm{a}_{\rm fin}}\dd\bm{a}(\pd/\pd\bm{a})E_{{\rm exp};\bm{a}}(x)$ and $\Delta F_{\rm exp}:=F_{\rm exp}(\bm{a}_{\rm fin})-F_{\rm exp}(\bm{a}_{\rm ini})$. This form of the second law is our third main result. 

	Remarkably, Eq.~\eqref{eq:second-law-work} is both experimentally accessible and ``gauge invariant''. Indeed, any equilibrium Markov embedding satisfying Assumption 3 and reproducing the same target dynamics and marginal equilibrium PDF gives $\sigma_{\rm tot}=\sigma_{\rm exp}$ for transitions between canonical states (see Methods and SI; cf. Ref.~\cite{DechantPRL2026} for the GLE class).

\paragraph{Zwanzig jump models.}
    \begin{figure*}
        \centering
        \includegraphics[width=175mm]{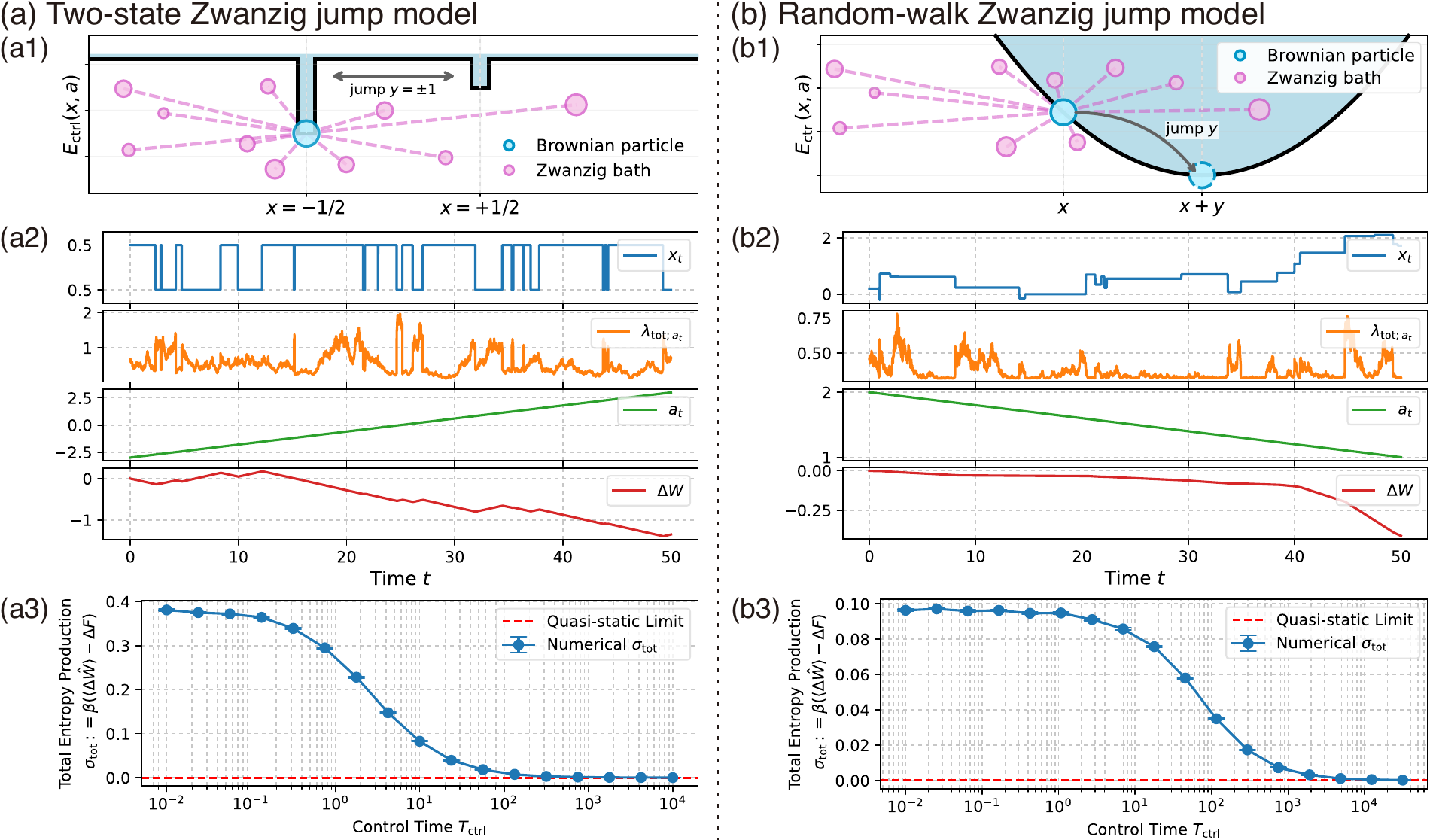}
        \caption{
            Two non-Markov jump models satisfying Assumptions 0--3 with $\phi(\tau)\propto e^{-|\tau|/\tau^*}$ and $\tau^*>0$. (a)~Two-state Zwanzig jump model~\eqref{eq:non-Markov-2-level}: (a1)~Schematic microscopic setup where the Brownian particle (blue) is coupled to the Zwanzig-bath particles (pink) via harmonic potentials (dashed pink line) for an even parity $\eps=+1$. (a2)~Sample paths of $(x_t, \lambda_{{\rm tot};a_t}, a_t,\Delta W)$ and (a3)~the second law~\eqref{eq:second-law-work} for the finite-time protocol with control time $T_{\rm ctrl}$ during the equilibrium-to-equilibrium transition with $a_t=a_{\rm ini}+(a_{\rm fin}-a_{\rm ini})t/T_{\rm ctrl}$. Panel~(b) shows the random-walk Zwanzig jump model~\eqref{eq:non-Markov-RW}: (b1) schematic microscopic setup, (b2) sample paths, and (b3) the second law. These two models belong to the Zwanzig jump family, and panels (a1) and (b1) illustrate their microscopic Zwanzig-type interpretation.
        }
        \label{fig:sim-nonMarkov-twolevel}
    \end{figure*}
	This naturally raises another question: can we construct non-Markov models that are both thermodynamically consistent and grounded in microscopic dynamics? To answer this question, we introduce the {\it Zwanzig jump models} as natural jump-process counterparts of the classical Zwanzig model~\cite{ZwanzigB}---the seminal microscopic oscillator-bath model originally introduced to derive the GLE (its quantum counterpart is the Caldeira--Leggett model). Our framework also encompasses the generalised Glauber dynamics for biological inference~\cite{ChenPRX2023} and allows its non-Gaussian extensions.

	With a positive function $M(s)$, we assume a quadratic auxiliary-field energy (called the {\it Zwanzig bath energy}): 
	\begin{equation}
			E_{\rm nctrl}[z,w]=\frac{1}{2}\int_0^\infty M(s)\left(w^2(s)+z^2(s)\right)\dd s
			\label{eq:energy-auxiliary-field-quadratic}
	\end{equation}
	For an Arrhenius-type transition rate, the history-dependent intensity functional is given by
	\begin{gather}
		\label{eq:def:Zwanzig-jump-model-intensity}
		\lambda_{\bm{a}}[y\mid X_t,\eta_t]=\lambda_0\rho_{\bm{a}}(y\mid x)e^{-\frac{\beta}{2}\Delta E_{\rm tot}} \\		
		\Delta E_{\rm tot} := \Delta E_{\rm ctrl}+y\int_{t_0}^t \phi(t-\tau)v_{\tau}\dd \tau + y\eta_t + \frac{y^2 \phi(0)}{2}\notag
	\end{gather}
	with $\Delta E_{\rm ctrl}:=E_{\rm ctrl}(x+y,\bm{a})-E_{\rm ctrl}(x,\bm{a})$, where $\rho_{\bm{a}}(y\mid x)=\rho_{\bm{a}^*}(\eps y\mid \eps x)$ and $\rho_{\bm{a}}(y\mid x)=\rho_{\bm{a}^*}(-\eps y\mid \eps x')$ with $y:=x'-x$ to satisfy Assumption 2. Here, $\phi(\tau):=\int_0^\infty M(s)\cos (s\tau)\dd s$ is the characteristic memory kernel. Also, $\eta_t:= \int_{0}^\infty M(s)w_{t}^{\rini}(s)\dd s$ is the coloured Gaussian noise originating from the bath initial condition $H_0$ and satisfies the fluctuation-dissipation relation: 
	\begin{equation}
		\la \eta_t\eta_{t+\tau}\ra_{\rm can} = \beta^{-1}\phi(\tau).
	\end{equation}
	We refer to these processes as Zwanzig jump models, which satisfy Assumptions 0--3 (see SI).

	We stress that the Zwanzig jump models are genuinely beyond the semi-Markov class: unlike a semi-Markov process, whose rate is fixed by the current state and the age since the last jump, Eq.~\eqref{eq:def:Zwanzig-jump-model-intensity} depends on the convolution memory of the entire jump history. This model is not covered by conventional semi-Markov stochastic thermodynamics~\cite{MaesSemiMarkov2009,JannPRX2022} but can be addressed in our formulation.

	Notably, our framework accommodates more general non-Markov models beyond the Zwanzig jump model. Indeed, we identify the Ginzburg--Landau memory-bath energy as an admissible non-Gaussian memory-bath energy (see Methods):
	\begin{align}
		\label{eq:gl-energy}			
		E_{\rm nctrl}:=& \int_{-\infty}^\infty \dd l \left[\kappa\left(\frac{\pd B_t(l)}{\pd l}\right)^2 + U(B_t(l))\right], \\
		B_t(l) :=& \int_{0}^\infty \dd s M(s)\left[w_t(s)\cos(sl)+z_t(s)\sin(sl)\right], \notag
	\end{align}
    where $\kappa$ is a non-negative constant and $U$ is a potential chosen such that the canonical PDF is normalisable. This energy functional has the standard form of a one-dimensional Euclidean classical field theory. Since $U$ can be non-quadratic, this example shows that the Fourier-embeddable class is not restricted to Gaussian Zwanzig baths but also includes intrinsically non-Gaussian memory-bath models.

\paragraph{Microscopic interpretation.}
	 The Zwanzig jump model can be regarded as a jump-process counterpart of the classical Zwanzig model when $\eps=+1$. Indeed, by introducing $q_t(s):=x_t-w_t(s)$ and $p_t(s):=(M(s)/s)z_t(s)$, we obtain the dynamical equations of infinitely many harmonic oscillators coupled to the target system $x_t$: 
	\begin{equation}
		\frac{\dd q_t(s)}{\dd t}=\frac{p_t(s)}{m(s)},\quad \frac{\dd p_t(s)}{\dd t}=-s^2m(s)(q_t(s)-x_t)
	\end{equation}
	with the ``mass'' $m(s):=M(s)/s^2$. The corresponding non-controllable energy is given by
	\begin{equation}
		E_{\rm nctrl} = \int_{0}^\infty \left[\frac{p_t^2(s)}{2m(s)}+\frac{s^2 m(s)}{2}(q_t(s)-x_t)^2\right]\dd s.
	\end{equation}
	Thus, this model family has a physically transparent microscopic interpretation. Within this model, we present two non-Markov examples that can be studied within our stochastic-thermodynamics framework. Microscopic interpretations of the two concrete examples are illustrated in Fig.~\ref{fig:sim-nonMarkov-twolevel}(a1) and (b1).

\paragraph{Example 1: two-state Zwanzig jump model.}
    The first example is the history-dependent two-state model, which we call the {\it two-state Zwanzig jump model}. It is characterised by a binary variable $x_t=\pm 1/2$. For a jump size $y=\pm 1$, the intensity functional depends on the history $X_t$ [Fig.~\ref{fig:sim-nonMarkov-twolevel}(a2) for its sample paths]:
    \begin{align}
        \lambda_{\bm{a}} [y\>|\> X_t, \eta_t] \!\propto\! \rho(y \>|\> x) e^{-\frac{y \beta}{2}
				\left\{
					-a
					+\int_{t_0}^t  \phi(t-\tau)v_{\tau}\dd \tau + \eta_t
                    \right\}
		}.
        \label{eq:non-Markov-2-level}
    \end{align}
	For $\eps=+1$, $x_t$ is the well index of a Brownian particle in a tilted double well [Fig.~\ref{fig:sim-nonMarkov-twolevel}(a1)], with an even control parameter $a^*=a$; for $\eps=-1$, the same model describes a classical spin in a magnetic field $a$ (odd parity: $a^*=-a$). In both cases $E_{{\rm ctrl};a}(x_t)=-ax_t$, and the flipping rate depends on the history of previous transitions via the memory kernel $\phi(\tau)$. The jump-size PDF is specified as $\rho(y\mid x)=\delta_{x,-1/2}\delta(y-1)+\delta_{x,1/2}\delta(y+1)$. See Fig.~\ref{fig:sim-nonMarkov-twolevel}(a3) for a numerical test of the second law $\sigma_{\rm tot} = \beta \left(\la \Delta W \ra -\Delta F\right) \geq 0$ in an equilibrium-to-equilibrium transition with the finite-time driving protocol. Even in this two-state case, the process is not semi-Markovian.

\paragraph{Example 2: random-walk Zwanzig jump model.}
    The second example is the history-dependent random-walk model for an even-parity variable $x^*=x$ (i.e., particle position) with $\eps=+1$. For a jump size $y$, the history-dependent intensity is given by 
    \begin{equation}
        \lambda_a[y \!\mid\! X_t, \!\eta_t] \!\propto\! \rho(y)e^{-\frac{\beta}{2}\left\{\Delta E_{\rm ctrl}+\frac{y^2}{2}\phi(0)+y\int_{t_0}^t \phi(t-\tau)v_{\tau}\dd \tau + y\eta_t\right\}}.
        \label{eq:non-Markov-RW}
    \end{equation}
	We refer to this model as the {\it random-walk Zwanzig jump model} in the harmonic potential $E_{{\rm ctrl};a}(x)=ax^2/2$ (even parity: $a^*=a$) with Gaussian jump-size distribution $\rho(y)=(1/\sqrt{2\pi \sigma^2})e^{-y^2/(2\sigma^2)}$ [see Fig.~\ref{fig:sim-nonMarkov-twolevel}(b1)]. Even though the jump-size distribution is Gaussian, the resulting process is not a Gaussian Langevin process and is not captured by the GLE framework. Just as in the two-state model, we allow the transition rate to depend on the history of previous transitions through the memory kernel $\phi(\tau)$. See Fig.~\ref{fig:sim-nonMarkov-twolevel}(b2) for its sample paths. Also, see Fig.~\ref{fig:sim-nonMarkov-twolevel}(b3) for a numerical test of the second law $\sigma_{\rm tot} = \beta \left(\la \Delta W \ra -\Delta F\right) \geq 0$ in an equilibrium-to-equilibrium transition.
    
\paragraph{Discussion.}
	In this work, we lift the long-standing Markov constraint and lay the foundations for non-Markov stochastic thermodynamics. Furthermore, we present the Zwanzig jump model as a minimal thermodynamically consistent model of history-dependent jump fluctuations beyond both GLEs and semi-Markov processes. With its transparent microscopic interpretation, this model provides a natural starting point for experimental and data-driven modelling of non-Markovian point processes, where memory affects transition intensities rather than continuous Langevin forces.
    
    A possible prototypical experimental setup is a colloidal bead hopping in an optically generated double-well potential within a passive viscoelastic medium. Tracer dynamics in such passive viscoelastic environments often exhibits anomalous diffusion~\cite{Szymanski2009,Wong2004} and, in viscoelastic actin networks, can be described by GLEs~\cite{Grebenkov2013}. Coarse-graining to the well index then yields a history-dependent two-state jump process. In the Kramers regime, bath relaxation after previous hopping events can affect the transition intensity, naturally leading to a Zwanzig-jump-type effective description. 
    
    Our theory offers a thermodynamic guiding principle for strong-memory processes: experimental models of history-dependent thermal fluctuations should satisfy the
	thermodynamic consistency conditions (Assumptions 1--2). One practical route to estimating the intensity functional is to reformulate existing machine-learning techniques for point processes~\cite{Omi2019} so as to incorporate Assumptions 1--2 and ensure consistency with thermodynamic laws. Such physics-informed, data-driven modelling will be an important direction for future work. 
    
    We also highlight several open questions concerning the Fourier embedding. A central question is how broad the class of Fourier-embeddable processes is. Our framework accommodates models with non-Gaussian bath statistics that are beyond the Zwanzig-type jump models. Systematically characterising this class, extending the framework to quantum systems, and clarifying its connection to the pseudomode framework are important open problems. Another fundamental question is whether the Fourier-embeddable class captures the full physically relevant range of thermodynamically consistent non-Markov jump processes, or whether important models lie beyond it.

\subsection*{Acknowledgements}
\begin{acknowledgments}
	KK was supported by JSPS KAKENHI (Grant Numbers 24H00833, 22H01141, and 23H00467) and JST PRESTO (Grant Number JPMJPR20M2). AD was supported by JSPS KAKENHI (Grant Numbers 22K13974, 24H00833, and 25K00926). The authors thank T. Sagawa, T. J. Kobayashi, H. Tajima, S. Ito, and S. Shimoe for fruitful discussions. In particular, the comments by M. L. Rosinberg and J. van der Meer were very helpful in improving our work.
\end{acknowledgments}

\subsection*{Competing interests}
	The authors declare no competing interests.

\subsection*{Author contributions}
	Both KK and AD contributed to the research conceptualisation and the analytical calculations: particularly, KK developed the Fourier embedding and derived thermodynamic laws, and AD co-developed the invariance of entropy production and its physical interpretation. Both KK and AD wrote the paper and confirmed all the findings. 

\section{Methods}
	We outline the derivations of our main results (see SI for more technical details).

\subsection{Derivation of the Markovian SPDE~\eqref{eq:set_complete_dynamics_hdCP}}
	Equation~\eqref{eq:set_complete_dynamics_hdCP} is equivalent to the second-order stochastic partial differential equation (SPDE)
	\begin{equation}
		\frac{\partial^2 z_t(s)}{\partial t^2} = -s^2z_t(s) + s v_{t}, 
	\end{equation}
	whose explicit solution is given by
    \begin{align*}
		z_t(s) =& c_0(s)\cos s(t-t_0) + c_1(s) \sin s(t-t_0)	\\    
			&+ \int_{t_0}^t  \dd \tau v_{\tau}\sin \{s(t-\tau)\}, \\
		w_t(s) =& c_1(s) \cos s(t-t_0) \!-\! c_0(s)\sin s(t-t_0) \\
		&+ \int_{t_0}^t  \dd \tau v_{\tau}\cos \{s(t-\tau)\}.
	\end{align*}
	The coefficients $c_0(s)$ and $c_1(s)$ are determined by the initial conditions, such that $z_{t_0}(s) = c_0(s)$ and $w_{t_0}(s) = c_1(s)$, to obtain Eq.~\eqref{def:Fourier_MarkovEmbedding_HDP}. Thus, the SDE~\eqref{eq:set_complete_dynamics_hdCP} is consistent with the Markovian-embedding representation~\eqref{def:Fourier_MarkovEmbedding_HDP}.

\subsection{Derivation of the fME~\eqref{eq:field_master_hdCP}}
	For an arbitrary functional $f[\Gamma_t]$, its time evolution $\dd f:= f[\Gamma_{t+\dd t}]-f[\Gamma_{t}]$ during $[t,t+\dd t)$ is given by 
	\begin{align}
		\dd f = \dd t
			\int_0^\infty \!\! s\left[w(s)\frac{\delta f}{\delta z(s)}-z(s)\frac{\delta f}{\delta w(s)}\right]\dd s +o(\dd t) 
	\end{align}
	without jumps with probability $1 - \dd t\lambda_{{\rm tot};\bm a}[\Gamma_t]$ or 
	\begin{equation}
		\dd f = f[\Gamma_t+\Delta \Gamma_{y}]-f[\Gamma_t] + o(\dd t^0)
	\end{equation}
	when a jump of size $y$ occurs with probability $\dd t \dd y \lambda_{\bm{a}}[y\mid \Gamma_t]$. Here we use $\left\la \dd f\right\ra = \dd t\int \dd \Gamma f[\Gamma](\pd P_{t}[\Gamma]/\pd t)$ and 
	\begin{equation}
	\begin{aligned}
		&\int \dd \Gamma P_t[\Gamma]\int_0^\infty \dd s \left[sw\frac{\delta f}{\delta z}-sz\frac{\delta f}{\delta w}\right] \\
		=& \int \dd \Gamma f[\Gamma]\int_0^\infty \dd s\left[-sw\frac{\delta P_t[\Gamma]}{\delta z}+sz\frac{\delta P_t[\Gamma]}{\delta w}\right],
	\end{aligned}
	\end{equation}
	by integration by parts. Let us define the total-system intensity $\Lambda_{\bm a}[\Gamma\mid \Gamma']:=\lambda_{\bm{a}}[x-x'\mid \Gamma']\delta[z-z']\delta [w-w'-(x-x')]$ with the $\delta$ functionals, $\delta[z-z']:=\prod_{s>0}\delta(z(s)-z'(s))$ and $\delta[w-w'-(x-x')]:=\prod_{s>0}\delta(w(s)-w'(s)-(x-x'))$. We then have 
	\begin{equation}
	\begin{aligned}
		&\int \dd \Gamma P_t[\Gamma]\int_{-\infty}^{\infty} \dd y\lambda_{\bm{a}}[y\mid \Gamma]f[\Gamma+\Delta \Gamma_{y}]  \\
		= &\int \dd \Gamma f[\Gamma]\int_{-\infty}^{\infty} \dd y\lambda_{\bm{a}}[y\mid \Gamma-\Delta \Gamma_{y}]P_t[\Gamma-\Delta \Gamma_{y}]  \\
		= &\int \dd \Gamma \dd \Gamma' f[\Gamma] \Lambda_{\bm{a}}[\Gamma \mid \Gamma']P_t[\Gamma']
	\end{aligned}
	\end{equation}
	by variable transformations $\Gamma+\Delta \Gamma_{y} \to \Gamma$ for the second line and $\Gamma-\Delta \Gamma_y \to \Gamma'$ for the third line with $\dd y\lambda_{\bm{a}}[y\mid \Gamma-\Delta \Gamma_{y}]=\dd \Gamma'\Lambda_{\bm a}[\Gamma\mid \Gamma']$. 
	We thus obtain 
	\begin{align}
		\int \dd \Gamma f[\Gamma]\frac{\pd}{\pd t}P_{t}[\Gamma] = \int \dd \Gamma f[\Gamma] \left(\mcL_{\rm A} + \mcL_{\rm J}\right)P_{t}[\Gamma].
	\end{align}
	Since this identity holds for an arbitrary functional $f[\Gamma]$, we obtain the fME~\eqref{eq:field_master_hdCP}.

	Note that the advective part of the fME is equal to that for harmonic oscillators and is a time-reversal-invariant Hamiltonian flow (see SI). The time-reversal symmetry is controlled entirely by the jump part of the fME.

\subsection*{Derivation of the second law~\eqref{eq:second-law}}
	By using the fME~\eqref{eq:field_master_hdCP}, we obtain 
	\begin{align}
		\dsgm_{\rm tot} 
		=& \int \dd \Gamma \dd \Gamma'\Lambda_{\bm{a}}[\Gamma'\mid \Gamma]P_t[\Gamma]\ln \frac{\Lambda_{\bm{a}}[\Gamma' \mid \Gamma]P_t[\Gamma]}{\Lambda_{\bm{a}^*}[\Gamma^*\mid \Gamma'^*]P_t[\Gamma']} \notag \\
		\geq& \int \!\! \left(\!1\!-\!\frac{\Lambda_{\bm{a}^*}[\Gamma^*\mid \Gamma'^*]P_t[\Gamma']}{\Lambda_{\bm{a}}[\Gamma' \mid \Gamma]P_t[\Gamma]}\!\right)\!\Lambda_{\bm{a}}[\Gamma'\mid \Gamma]P_t[\Gamma]\dd \Gamma \dd \Gamma' \notag \\
		=& \int \!\!\lambda_{{\rm tot}; \bm{a}}[\Gamma]P_t[\Gamma]\dd \Gamma \!\!-\!\! \int \!\!\lambda_{{\rm tot}; \bm{a}^*}[\Gamma^*]P_t[\Gamma]\dd \Gamma \!\!=\!\! 0,
	\end{align}
	where we used the following relations: $\ln(t^{-1})\geq 1-t$ for all $t>0$, $\lambda_{{\rm tot}; \bm{a}}[\Gamma] = \lambda_{{\rm tot}; \bm{a}^*}[\Gamma^*]$, and 
	\begin{equation}
	\begin{aligned}
		&\int  \ln P_t[\Gamma]\left(sz\frac{\delta}{\delta w}-sw\frac{\delta}{\delta z}\right)P_t[\Gamma]\dd s\dd \Gamma \\
		=& -\int  \left(sz\frac{\delta}{\delta w}-sw\frac{\delta}{\delta z}\right)P_t[\Gamma]\dd s\dd \Gamma = 0
	\end{aligned}
	\end{equation}
	by integration by parts.

\subsection*{General solution of the fME~\eqref{eq:field_master_hdCP}}
    Under Assumptions 1 and 2, the stationary condition of the fME reduces to $\mcL_{\rm A}P_{\mrcan;\bm a}=0$, which can be solved by the method of characteristics. The resulting stationary canonical distribution imposes the following necessary form on the energy functional:
	\begin{equation}
	\begin{gathered}
		\label{eq:fME-gen_sol}
		E_{\bm{a}}[\Gamma] \!=\! \mathcal{G}_{\bm{a}}\left(x, \{r(s),\psi(s)\}_s\right), \>\>
		z(s) \!:=\! r(s)\sin\theta(s), \\
		w(s) \!:=\! r(s)\cos\theta(s), \>\>
		\psi(s) \!:=\! \theta(s)-\frac{s}{s^*}\theta(s^*),
	\end{gathered}
	\end{equation}
	where $s^*>0$ is an arbitrary reference wave number and $\mathcal{G}_{\bm{a}}$ is an arbitrary functional such that the canonical distribution is normalisable and satisfies the periodic boundary condition for the polar coordinates:
	\begin{align}
		&\mathcal{G}_{\bm{a}}\left(x, \{r(s),\psi(s)\}_s\right) \label{eq:fME-gen_sol_PBC} \\
		= &\mathcal{G}_{\bm{a}}\left(x, \left\{r(s),\psi(s)+2\pi \left[n(s)-\frac{s}{s^*}n(s^*)\right]\right\}_s\right)\notag
	\end{align}
	for any integer function $n(s)$.

	\paragraph{Derivation.} Under Assumptions 1 and 2, we obtain
	\begin{align}
		\mcL_{\rm J}P_{\mrcan;\bm{a}}[\Gamma] \!=\! \left(\lambda_{{\rm tot};\bm{a}^*}[\Gamma^*]\!-\!\lambda_{{\rm tot};\bm{a}}[\Gamma]\right)P_{\mrcan;\bm{a}}[\Gamma] \!=\! 0,
	\end{align}
	where we used the invariance under time reversal for the volume element $\dd \Gamma'=\dd \Gamma'^*$ and the total intensity $\lambda_{{\rm tot};\bm{a}^*}[\Gamma^*]=\lambda_{{\rm tot};\bm{a}}[\Gamma]$. Therefore, it is enough to solve
	$\mcL_{\rm A}P_{\mrcan;\bm{a}}[\Gamma]  =  0$ in its steady state with the control parameter $\bm{a}$ fixed. This equation is a first-order functional differential equation and can be solved by the method of characteristics. Let us introduce the polar coordinate
	\begin{equation}
		z(s)=r(s)\sin\theta(s),\quad w(s)=r(s)\cos\theta(s).
	\end{equation}
	Then, the advective equation reduces to
	\begin{gather}
		\mcL_{\rm A}P_{\mrcan;\bm{a}}=-\int_0^\infty \dd s\,s\frac{\delta P_{\mrcan;\bm{a}}}{\delta \theta(s)} = 0 \\
		P_{\mrcan;\bm{a}}[x,\{r(s),\theta(s)\}]=P_{\mrcan;\bm{a}}[x,\{r(s),\theta(s)+2\pi n(s)\}] \notag
	\end{gather}
	for any integer-valued function $n(s)$. This condition is necessary and sufficient for the polar-coordinate representation to define a global single-valued distribution. The Lagrange-Charpit equations are given by
	\begin{equation}
	\begin{aligned}
		&\frac{\dd r_\mu(s)}{\dd \mu} = 0,\quad
		\frac{\dd \theta_\mu(s)}{\dd \mu} = -s,\quad
		\frac{\dd P_\mu}{\dd \mu}=0 \\
		\Longrightarrow \quad
		&r_\mu(s)=A(s),\quad \theta_\mu(s)=-s\mu+B(s)
	\end{aligned}
	\end{equation}
	for any $s>0$, where $\mu$ is the parameter for the method of characteristics and $A(s),B(s)$ are constants along each characteristic. By rewriting
	\begin{equation}
		C(s):=B(s)-\frac{s}{s^*}B(s^*)=\theta_\mu(s)-\frac{s}{s^*}\theta_\mu(s^*)
	\end{equation}
	with a positive reference wave number $s^*>0$, the general solution is given by
	\begin{equation}
		\label{eq:der-gen-sol-fME}
		P_{\mrcan;\bm{a}}[\Gamma]
		\!=\!g_{\bm{a}}\left(x,\{r(s),\psi(s)\}_s\right),\>\>
		\psi(s)\!:=\!\theta(s)\!-\!\frac{s}{s^*}\theta(s^*),
	\end{equation}
	with an arbitrary functional $g_{\bm{a}}$ satisfying the corresponding periodic boundary condition. We thus obtain the most general stationary canonical distribution compatible with $\mcL_{\rm A}P_{\mrcan;\bm a}=0$ under Assumptions 1 and 2. Finally, the periodic boundary condition in $\theta(s)$ is equivalent to Eq.~\eqref{eq:fME-gen_sol_PBC}.
	
	Note that the Zwanzig bath energy~\eqref{eq:energy-auxiliary-field-quadratic} satisfies this condition. The Ginzburg--Landau memory-bath energy~\eqref{eq:gl-energy} provides another admissible class of energy functionals as shown below.

\subsection*{Gauge invariance of cumulative entropy production}
    Let us prove the gauge invariance of the cumulative entropy production by expressing it entirely in terms of the driven target dynamics and the marginal equilibrium PDF $P_{{\rm can};\bm{a}}(x)$. Let $\chi$ represent the auxiliary variables of an arbitrary equilibrium Markov embedding, such that $\Gamma=(x,\chi)$. Under Assumption 3, $P_{{\rm can};\bm{a}}(x)=e^{\beta[F(\bm{a})-E_{{\rm ctrl};\bm{a}}(x)-E'_{\rm nctrl}(x)]}$, where the coarse-grained non-controllable energy $E'_{\rm nctrl}(x)$, defined by $e^{-\beta E'_{\rm nctrl}(x)}:=\int\dd\chi\,e^{-\beta E_{\rm nctrl}(x,\chi)}$, is independent of $\bm{a}$. By comparing this expression with the definitions of $E_{{\rm exp};\bm{a}}$ and $F_{\rm exp}$, we obtain $E_{{\rm exp};\bm{a}}(x)=E_{{\rm ctrl};\bm{a}}(x)+E'_{\rm nctrl}(x)+c(\bm{a})$ and $F_{\rm exp}(\bm{a})=F(\bm{a})+c(\bm{a})$, where $c(\bm{a})$ is an $x$-independent function. We therefore obtain $\Delta W_{\rm exp}=\Delta W+\Delta c$ and $\Delta F_{\rm exp}=\Delta F+\Delta c$, proving Eq.~\eqref{eq:second-law-work}. Equivalently, $\sigma_{\rm exp}=-\la\int\dd t\,\dot{\bm{a}}_t(\pd/\pd\bm{a}_t)\ln P_{{\rm can};\bm{a}_t}(x_t)\ra$, which contains only the target-system statistics. Thus, all equilibrium Markov embeddings satisfying Assumption 3 and reproducing the same target dynamics and marginal equilibrium PDF give the same cumulative entropy production. This establishes gauge invariance with respect to the choice of embedding.

\subsection*{Beyond the Zwanzig jump models}
	Our framework is not restricted to the quadratic Zwanzig bath energy. As a simple example, we introduce the Ginzburg--Landau memory-bath energy:
	\begin{gather}	
		E_{\rm nctrl}:= \int_{-\infty}^\infty \dd l \left[\kappa\left(\frac{\pd B_t}{\pd l}\right)^2 + U(B_t)\right], \\
		B_t(l) := \int_{0}^\infty \dd s M(s)\left[w_t(s)\cos(sl)+z_t(s)\sin(sl)\right], \notag
	\end{gather}
	with a non-negative constant $\kappa\geq 0$. The local potential $U$ is chosen such that the canonical PDF $P_{{\rm can};\bm{a}}[\Gamma]$ is normalisable and the energy is invariant under time reversal. This construction can be understood as the thermodynamic-limit form of a finite-mode regularization with periodic boundary conditions (see SI). The field $B_t(l)$ can be decomposed into the following form:
	\begin{gather}
		B_t(l) = \eta_{t-l} + \int_{t_0}^t \phi(t-\tau-l)v_{\tau}\dd \tau, \\
		\phi(\tau) := \int_{0}^\infty \!\!M(s)\cos(s\tau)\dd s, \quad
		\eta_{t} := \int_0^\infty \!\! M(s)w_{t}^{\rini}(s)\dd s.\notag 
	\end{gather}
	For a non-quadratic choice of $U$, the equilibrium distribution of the auxiliary field is generally non-Gaussian. Consequently, $\eta_t$ is a coloured and non-Gaussian stationary process originating from the bath initial condition.

\end{document}


\title{	
				Supplementary Information: \\Stochastic thermodynamics for classical non-Markov jump processes
			}
\author{Kiyoshi Kanazawa}
\author{Andreas Dechant}
\affiliation{
			Department of Physics, Graduate School of Science, Kyoto University, Kyoto 606-8502, Japan
		}
\date{August 4, 2026}
\pacs{}

\maketitle
\section{Mathematical notation}\label{sec:notation}
	
	\subsection{General mathematical notation}
		The set $\mathbb{R}$ represents the set of real numbers. The set of positive real numbers is written as $\mathbb{R}^+:= \{s \mid s > 0, s\in \mathbb{R}\}$. In this article, $s$ typically represents the frequency in the Fourier space. We sometimes use the simple notation for sets with frequency index, such that $\{z(s)\}_{s} := \{z(s) \mid s \in \mathbb{R}^+\}$.

		Indicator functions are denoted by $\mathbb{I}_{A}$: $\mathbb{I}_{A}=1$ if the statement $A$ is true; otherwise $\mathbb{I}_{A}=0$ (e.g., $\mathbb{I}_{x=1}=1$ for $x=1$ and $\mathbb{I}_{y=0}=0$ for $y=1$). 

	\subsection{Notation for stochastic variables}
		In this report, we emphasize any stochastic variable by putting the hat symbol, such that $\hat{A}$. Without the hat symbol, the variable is just a real number (e.g., $A$). The probability density function (PDF) is written as $P_t(A):=P(\hat{A}_t=A)$, which characterizes the probability $\hat{A}_t\in [A,A+\dd A)$ by $P_t(A)\dd A$. Correspondingly, the ensemble average of any stochastic variable $\hat{A}$ is denoted by $\la\hat{A}\ra:=\int AP_t(A)\dd A$.	Using this notation, with Dirac's $\delta$ function $\delta(A-A')$, the PDF can be rewritten as the following identity
		\begin{equation}
			P_t(A) = \int \delta(A-A')P_t(A')\dd A'=\la \delta(A-\hat{A}_t)\ra.
			\label{eq:notation:PDFtodeltafunc}
		\end{equation}

	\subsection{Notation for functionals}
		In this article, {\it functionals}---maps whose arguments are functions, such as $f[\{z(s)\}_s]$ with a function $\{z(s)\}_s$---frequently appear. In our notation, we write functionals with square brackets to stress that the map is a functional. Functionals are abbreviated as $f[z]$ when their meaning is evident. 

		Using functionals, we can describe the stochastic properties of any stochastic field variable $\{\hz_t(s)\}_s$. For example,  the probability that $\hz_t(s) \in \prod_s [z(s),z(s)+\dd z(s))$ is given by $P_t[z]\mathcal{D}z$ with the probability density functional $P_t[z]$, which is called the PDF for short, and the functional volume element $\mathcal{D}z:=\prod_s \dd z(s)$. The ensemble average of any functional $f[\hz_t]$ is given by the path integral, such that 
		\begin{equation}
			\la f[\hz]\ra := \int f[z]P_t[z]\mcD z =  \int f[z]P_t[z] \left(\prod_s \dd z(s)\right). 
			\label{eq:def:ensemble_average_functional}
		\end{equation}
		Similarly to Eq.~\eqref{eq:notation:PDFtodeltafunc}, the PDF can be written as the average of the $\delta$ functional, 
		\begin{equation}
			P_t[z] = \int \delta[z-z']P_t[z']\mathcal{D}z' =\la \delta[z-\hz_t]\ra, \>\>\> 
			\delta[z-\hz_t]:=\prod_s \delta(z(s)-\hz_t(s)).
		\end{equation}

\section{Model and field master equation}\label{sec:model}
	\subsection{History-dependent jump process with bath initial condition}
		\begin{figure*}
			\centering
			\includegraphics[width=90mm]{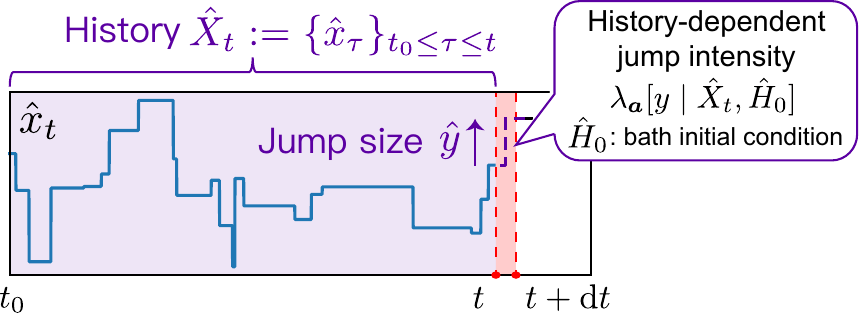}
			\caption{
				Schematic of the one-dimensional non-Markov jump process (or the history-dependent jump process) with the bath initial condition $\hH_0:=\{\hz_{t_0}(s),\hw_{t_0}(s)\}_{s>0}$. The jump intensity density $\lambda_{\bm a}[y\mid \hX_t,\hH_0]$ depends on the history, where $\hX_t:=\{\hx_{\tau}\}_{t_0\leq \tau\leq t}$ represents the entire history, and $\bm{a}$ is the external control parameter, such as the potential modulation. The initial state $(\hx_{t_0},\hH_0)$ is sampled from the canonical distribution.
			}
		\end{figure*}
		In this article, we address a history-dependent jump process with the bath initial condition $\hH_0$:
		\begin{box_summary}{History-dependent jump process with the bath initial condition}
			\vspace{-5mm}
			\begin{equation}\label{eq:def_history_dependent_Poisson}
				\frac{\dd \hx_t}{\dt} = \hxiCP_{\lambda_{\bm{a}}[y \mid \hX_t,\hH_0]}, \>\>\>
				\hX_t := \{\hx_{\tau}\}_{t_0\leq \tau\leq t},\quad 
				\hH_0:=\{\hz_{t_0}(s),\hw_{t_0}(s)\}_{s>0},
				\vspace{-1mm}
			\end{equation}
		\end{box_summary}
		\noindent
		The bath initial condition $\hH_0$ and its time evolution are specified by Eqs.~\eqref{eq:init-condition-H0} and~\eqref{eq:set_complete_dynamics_initCondition}, respectively. Thermodynamic consistency requires the initial state $(\hx_{t_0},\hH_0)$ to be sampled from the canonical distribution~\eqref{eq:def:equilibrium-dist}. Also, $\lambda_{\bm{a}}[y\mid \hX_t,\hH_0]$ is the intensity density conditional on the full history of the system $\hX_t=\{\hx_{\tau}\}_{t_0\leq \tau\leq t}$ and the bath initial condition $\hH_0$, where $y$ is the jump size and $\bm{a}$ is an external control parameter. The velocity includes the impulses described by the Dirac $\delta$ functions due to jumps. 
		
		More technically, for any given history $\hX_t = \{\hx_{\tau}\}_{t_0\leq \tau\leq t}$ and with $\dd \hx_t := \hx_{t+\dt} - \hx_t$, Eq.~\eqref{eq:def_history_dependent_Poisson} implies that
		\begin{equation}
			\dd \hx_t := 	\begin{cases}
							\hat{y} & (\mbox{prob.} = \dt \dy\lambda_{\bm{a}}[y\mid \hX_t,\hH_0] \mbox{ for any } \hat{y} \in [y,y+\dy)) \\
							0 & (\mbox{prob.} = 1 - \dt \lambda_{{\rm tot};\bm{a}}[\hX_t,\hH_0])
						\end{cases}, \quad 
            \lambda_{{\rm tot};\bm{a}}[\hX_t,\hH_0]:=\int_{-\infty}^\infty \dy\lambda_{\bm{a}}[y\mid \hX_t,\hH_0].
		\end{equation}
		If the intensity $\lambda_{\bm a}$ does not depend on the history $\hX_t$ and the hidden variables $\hH_0$, this process reduces to the Markov jump process, a standard class studied in conventional stochastic thermodynamics.

	\subsection{Markovian embedding}
			\subsubsection{A brief history of Markov embedding}
			Our work is based on the Markov-embedding technique: a low-dimensional non-Markov system is converted into a higher-dimensional Markov system by adding auxiliary variables. In classical statistical physics, this technique has a long history in studies of the generalised Langevin equation (GLE)~\cite{Klippenstein2021} and has its roots in Mori's pioneering works~\cite{Mori1965,Mori1965CF} and early oscillator-bath models~\cite{Magalinskii1959,FordKacMazur1965,Zwanzig1973}. In particular, Zwanzig later used such a construction to derive exact nonlinear GLEs. Later, this technique was further developed, in particular for numerical computations~\cite{FerrarioGrigolini1979,ErmakBuckholz1980,CiccottiRyckaert1980}. Later, closely related ideas have also been developed in open quantum systems. Indeed, pseudomode methods~\cite{Imamoglu1994,Garraway1997} and the reaction-coordinate mapping~\cite{IlesSmith2014,Strasberg2016} are established techniques that introduce extended systems, both formulated for systems linearly coupled to Gaussian baths.

			The Markov-embedding technique has also been widely used beyond the GLE. For example, it has recently been applied to general non-Markov point processes~\cite{KzDidier2019PRL,KzDidier2021PRL,KzDidier2024PRR}.
			A closely related embedding technique was used to construct the generalised Glauber dynamics for classical discrete-state systems~\cite{ChenPRX2023}. This model, developed to infer the social dynamics of mice, couples Ising and Potts variables to harmonic-oscillator baths to generate reversible history-dependent transition rates.
			Unfortunately, there is no fixed terminology for this technique; terms such as Markov embedding, Markovianisation, and auxiliary-variable methods are used. In mathematical finance, the same general idea was further developed for stochastic Volterra processes around 2018 and is now often called Markovian lifting~\cite{CuchieroTeichmann2020}. 

		\subsubsection{Embedding of the path $\hX_t=\{\hx_{\tau}\}_{t_0\leq \tau\leq t}$}
			Before proceeding with the derivation of the corresponding fME, let us introduce a useful complete set of the system variables for the Markovian embedding. In the previous section, we used $(\hX_t, \hH_0)$ as a naive complete set of the system variables. However, the master equation is unknown for this non-Markov model under this variable representation, which makes it difficult to develop stochastic thermodynamics for the non-Markov jump processes. To solve this problem, we introduce the Fourier convolution transform of the velocity, such that

			\begin{box_summary}{Fourier embedding: $\hX_t=\{\hx_{\tau}\}_{t_0\leq \tau\leq t} \to \hGamma_t=(\hx_t,\{\hz_t(s),\hw_t(s)\}_{s> 0})$}
				\vspace{-5mm}
				\begin{equation}
					\label{def:Fourier_MarkovEmbedding_HDP}
					\hw_t(s) := \hw_t^{\rini}(s) + \int_{t_0}^t \hv_{\tau}\cos\{s(t-\tau)\}\dd \tau, \>\>\>
					\hz_t(s) := \hz_t^{\rini}(s) + \int_{t_0}^t \hv_{\tau}\sin\{s(t-\tau)\}\dd \tau, \>\>\>
					\hv_{\tau} := \frac{\dd \hx_{\tau}}{\dd \tau},
					\vspace{-1mm}
				\end{equation}
			\end{box_summary}
			\noindent
				which is defined for $s>0$. Here, $\{\hw_t^{\rini}(s),\hz_t^{\rini}(s)\}_{s>0}$, defined by Eq.~\eqref{eq:set_complete_dynamics_initCondition}, are the dynamical contributions originating from the bath initial condition $\hH_0$ defined by Eq.~\eqref{eq:init-condition-H0}. We then adopt the variable set
			\begin{equation}
				\hGamma_t := (\hx_t, \{\hz_t(s),\hw_t(s)\}_{s>0}) 
			\end{equation}
			as a useful complete variable set for the following formulation. 
				In this work, a history-dependent jump process is called {\it Fourier-embeddable} if its intensity is completely characterised by the Fourier representation:
			\begin{box_summary}{Fourier-embeddable history-dependent jump processes}
				\vspace{-5mm}
				\begin{equation}
					 \lambda_{\bm{a}}[y\mid \hX_t,\hH_0]=\lambda_{\bm{a}}[y\mid \hGamma_t]. 
					\vspace{-1mm}
				\end{equation}
			\end{box_summary}\noindent

		\subsubsection{Phase space}
			We regard the state variables $\hGamma_t$ as the point in the phase space $\mathbb{S}$, such that 
			\begin{equation}
				\hGamma_t \in \mathbb{S}, \>\>\> \mathbb{S}:= \{(x, \{z(s),w(s)\}_{s}) : x \in \mathbb{R}, \{z(s)\}_{s} \in \bm{F}, \{w(s)\}_s \in \bm{F} \}
			\end{equation}
			with the real number space $\mathbb{R}$, and the function space $\bm{F}$. 

	\subsection{Markovian field dynamics}
		For a Fourier-embeddable history-dependent process, the compound Poisson dynamics~\eqref{eq:def_history_dependent_Poisson} of the original variables $\{\hx_\tau\}_{t_0\leq \tau\leq t}$ is equivalent to the following SDE and SPDEs for the embedded variables $\hGamma_t:=(\hx_t,\{\hz_t(s),\hw_t(s)\}_{s>0})$:
		\begin{subequations}\label{eq:set_complete_dynamics}
			\begin{box_summary}{The SDE equivalent to the Fourier embedding}
				\vspace{-5mm}
				\begin{align}\label{eq:set_complete_dynamics_hdCP}
					\frac{\dd \hx_t}{\dt} = \hxiCP_{\lambda_{\bm{a}}[y\mid \hGamma_t]}, \>\>\> 
					\frac{\partial \hz_t(s)}{\partial t} = s\hw_t(s),\>\>\>
					\frac{\partial \hw_t(s)}{\partial t} = -s\hz_t(s) + \hxiCP_{\lambda_{\bm{a}}[y\mid \hGamma_t]}, 
					\vspace{-1mm}
				\end{align}
			\end{box_summary}\noindent
			where the jump term $\hxiCP_{\lambda_{\bm{a}}[y \mid \hGamma_t]}$ simultaneously acts on $\hx_t$ and $\hw_t(s)$ for all $s>0$. Indeed, the solution of Eq.~\eqref{eq:set_complete_dynamics_hdCP} is given by the following formula: 
			\begin{box_summary}{Solution of the SDE~\eqref{eq:set_complete_dynamics_hdCP}}
				\vspace{-5mm}
	                \begin{gather}\label{eq:set_complete_dynamics_initCondition}
					\hz_t(s) = \hz_t^{\rini}(s) + \int_{t_0}^t \hv_{\tau}\sin\{s(t-\tau)\}\dd \tau, \>\>\> 	
					\hw_t(s) = \hw_t^{\rini}(s) + \int_{t_0}^t \hv_{\tau}\cos\{s(t-\tau)\}\dd \tau , \>\>\>
					\hv_\tau := \frac{\dd \hx_\tau}{\dd \tau}, \\
					\hz_t^{\rm ini}(s):= \hz_{t_0}(s)\cos\{s(t-t_0)\} + \hw_{t_0}(s)\sin \{s(t-t_0)\}, \quad 
					\hw_t^{\rm ini}(s):= -\hz_{t_0}(s)\sin\{s(t-t_0)\} + \hw_{t_0}(s)\cos \{s(t-t_0)\}, \notag
					\vspace{-1mm}                    
				\end{gather}
			\end{box_summary}\noindent
			where $\hz_t^{\rini}(s)$ and $\hw_t^{\rini}(s)$ represent the dynamical components originating from the bath initial condition defined by
			\begin{box_summary}{Definition: the bath initial condition for the Fourier-embeddable jump process}
				\vspace{-5mm}
					\begin{equation}\label{eq:init-condition-H0}
					\hH_0:=\{\hz_{t_0}(s), \hw_{t_0}(s)\}_{s>0}. 
				\end{equation}
			\end{box_summary}\noindent		
		\end{subequations}
		Thus, the role of the bath initial condition $\hH_0$ is fixed at this stage. The joint initial state $(\hx_{t_0},\hH_0)$ is sampled from the canonical distribution, as defined by Eq.~\eqref{eq:def:equilibrium-dist}. This set of SDEs characterizes the complete dynamics of the phase point $\hGamma_t=(\hx_t, \{\hz_t(s),\hw_t(s)\}_{s})$ in a closed form.

	\subsection{Field master equation for the history-dependent Poisson process}
		The field master equation (fME) corresponding to the SPDEs~\eqref{eq:set_complete_dynamics_hdCP} is given by the following functional ME: 
		\begin{subequations}
			\begin{box_summary}{Field master equation after the Fourier embedding}
				\vspace{-5mm}
				\begin{equation}\label{eq:field_master_hdCP}
					\frac{\partial P_t[\Gamma]}{\partial t} = \left(\mcL_{\rm A}+\mcL_{\rm J}\right)P_t[\Gamma],
					\vspace{-1mm}
				\end{equation}
			\end{box_summary}
			\noindent
			where the advective and jump Liouville operators are defined by 
			\begin{align}
				\mcL_{\rm A}P_t[\Gamma] &:= 
				\int_0^\infty \ds \left\{s z(s)\frac{\delta P_t[\Gamma]}{\delta w(s)} - s w(s)\frac{\delta P_t[\Gamma]}{\delta z(s)}\right\} , \\
				\mcL_{\rm J}P_t[\Gamma] &:= \int_{-\infty}^{\infty}\dy\left\{\lambda_{\bm{a}}[y\mid \Gamma-\Delta \Gamma_y]P_t[\Gamma-\Delta \Gamma_y]-\lambda_{\bm{a}}[y\mid \Gamma]P_t[\Gamma]\right\} 
				= \int \dd \Gamma' \left\{\Lambda_{\bm{a}}[\Gamma \mid \Gamma']P_t[\Gamma']-\Lambda_{\bm{a}}[\Gamma' \mid \Gamma]P_t[\Gamma]\right\}, \notag \\
				\Lambda_{\bm{a}} [\Gamma \mid \Gamma'] &:= \int_{-\infty}^\infty \dd y \lambda_{\bm{a}}[y\mid\Gamma-\Delta \Gamma_y]\delta [\Gamma-\Delta \Gamma_y -\Gamma'] 
				= \lambda_{\bm{a}} [x-x'\mid \Gamma']\delta[z-z']\delta[w-w'-(x-x')],\notag
			\end{align}
		\end{subequations}
		where $\Delta \Gamma_y := (y, \{0 \}_s, \{y\}_{s})$ is the jump-size vector, and $\dd \Gamma := \dd x \mcD z \mcD w$ is the volume element. Here $P_t[\Gamma]$ denotes the probability density functional on the (infinite-dimensional) embedded phase space $\mathbb{S}$. Functional integrals such as $\int P_t[\Gamma](\dots)\dd \Gamma$
        should be understood in the standard field-theoretic sense as the formal continuum limit of a finite-dimensional (finite-$K$) regularization; see Sec.~\ref{sec:discussion:Rigor}.
		Here, $\lambda_{\bm{a}}[y\mid\Gamma]$ denotes the jump-size intensity of the target variable, whereas $\Lambda_{\bm{a}}[\Gamma'\mid\Gamma]$ denotes the induced transition-rate density on the embedded phase space.

		\subsubsection*{Derivation}
			For an arbitrary functional $f[\hGamma_t]$, its time evolution $\dd f:= f[\hGamma_{t+\dd t}]-f[\hGamma_{t}]$ during $[t,t+\dd t)$ is given by 
			\begin{align}
				\dd f = \begin{cases}
					\displaystyle
					\dd t
					\int_0^\infty \dd s\left[s \hw_t(s)\frac{\delta f}{\delta \hz_t(s)}-s \hz_t(s)\frac{\delta f}{\delta \hw_t(s)}\right] + o(\dd t) & (\mbox{no jumps; prob.}=1 - \dd t\lambda_{{\rm tot};\bm a}[\hGamma_t])\\
					f[\hGamma_t+\Delta \hGamma_{\hy}]-f[\hGamma_t] + o(\dd t^0) & (\mbox{jump size $\hy \in [y,y+\dd y)$; prob.}=\dd t\dd y\lambda_{\bm a}[y\mid \hGamma_t])
				\end{cases},
			\end{align}
			where we consider only the leading-order contribution and we define the total intensity $\lambda_{{\rm tot};\bm{a}}[\Gamma]:= \int_{-\infty}^{\infty} \dd y \lambda_{\bm a}[y\mid \Gamma]$. Note that we apply the Taylor functional expansion to the leading order. Thus, by taking the ensemble averages of both sides, we obtain 
			\begin{align}
				\left\la \dd f\right\ra =& \dd t \int \dd \Gamma P_t[\Gamma]\left\{
					\int_0^\infty \dd s\left(sw(s)\frac{\delta f}{\delta z(s)}-sz(s)\frac{\delta f}{\delta w(s)}\right) + \int_{-\infty}^\infty \dd y \lambda_{\bm a}[y \mid \Gamma](f[\Gamma+\Delta \Gamma_{y}]-f[\Gamma])
				\right\} + o(\dd t).
			\end{align}
			Here we use the following relations: 
			\begin{gather}
				\left\la \dd f\right\ra = \left\la f[\hGamma_{t+\dd t}]\right\ra - \left\la f[\hGamma_t]\right\ra = \int \dd \Gamma f[\Gamma]\left(P_{t+\dd t}[\Gamma]-P_t[\Gamma]\right) = \dd t\int \dd \Gamma f[\Gamma]\frac{\pd}{\pd t}P_{t}[\Gamma] + o(\dd t), \\
				\int \dd \Gamma P_t[\Gamma]\int_0^\infty \dd s \left(sw(s)\frac{\delta f}{\delta z(s)}- sz(s)\frac{\delta f}{\delta w(s)}\right)
				= \int \dd \Gamma f[\Gamma]\int_0^\infty \dd s\left(-sw(s)\frac{\delta P_t[\Gamma]}{\delta z(s)}+sz(s)\frac{\delta P_t[\Gamma]}{\delta w(s)}\right),
			\end{gather}
			by integration by parts, and 
			\begin{align}
				\int \dd \Gamma P_t[\Gamma]\int_{-\infty}^{\infty}\dd y \lambda_{\bm{a}}[y\mid \Gamma](f[\Gamma+\Delta \Gamma_{y}]-f[\Gamma]) = \int \dd \Gamma f[\Gamma]\int_{-\infty}^{\infty} \dd y(\lambda_{\bm{a}}[y\mid \Gamma-\Delta \Gamma_{y}]P_t[\Gamma-\Delta \Gamma_{y}]-\lambda_{\bm{a}}[y\mid \Gamma]P_t[\Gamma]),
			\end{align}
			by a variable transformation $\Gamma+\Delta \Gamma_{y} \to \Gamma$ for the first term. 
			We thus obtain
			\begin{equation}
			\begin{aligned}
				\int \dd \Gamma f[\Gamma]\frac{\pd}{\pd t}P_{t}[\Gamma] = \int \dd \Gamma f[\Gamma] \bigg [& \int_0^\infty \dd s\left(-sw(s)\frac{\delta P_t[\Gamma]}{\delta z(s)}+sz(s)\frac{\delta P_t[\Gamma]}{\delta w(s)}\right)\\
				&+\int_{-\infty}^{\infty} \dd y(\lambda_{\bm{a}}[y\mid \Gamma-\Delta \Gamma_{y}]P_t[\Gamma-\Delta \Gamma_{y}]-\lambda_{\bm{a}}[y\mid \Gamma]P_t[\Gamma]) \bigg].
			\end{aligned}
			\end{equation} 
			Since this identity holds for an arbitrary functional $f[\Gamma]$, we obtain the fME~\eqref{eq:field_master_hdCP}.

\section{Time-reversal symmetry for the non-Markov jump processes}\label{sec:time-reversal-symmetry}	
	\subsection{State-level time-reversal operation}
		Let us formulate the time-reversal operation on the level of the instantaneous state. The target system $\hx_{t}$ is assumed to have the parity of $\eps=\pm 1$. Also, the parities of $\hz_t$ and $\hw_t$ are given by $-\eps$ and $\eps$, respectively. In other words, the time-reversal operation of the instantaneous state $\hGamma_t$ based on the Fourier representation is given as follows: 
			\begin{box_summary}{Time-reversal operation of the instantaneous state (Fourier representation)}
				\vspace{-5mm}
				\begin{equation}
					\hGamma_{t}^* := (\hx^{*}_t, \{\hz^*_t(s),\hw^*_t(s)\}_{s>0}),\quad 
					\hx^{*}_t = \eps \hx_t, \>\>\> \hz^*_t(s) = -\eps \hz_t(s), \>\>\> \hw^*_t(s) = \eps \hw_t(s).
					\vspace{-1mm}
				\end{equation}
			\end{box_summary}\noindent
			Also, the control parameter $\bm{a}$ has its corresponding time reversal $\bm{a}^*$. 

	\subsection{Assumption 0: the existence of the stationary distribution in the field master equation}
		In the following, we assume that the fME~\eqref{eq:field_master_hdCP} has the stationary solution: 
		\begin{box_summary}{Assumption 0: Existence of the stationary solution}
			\vspace{-5mm}	
			\begin{equation}
				\left(\mcL_{\rm A}+\mcL_{\rm J}\right)P_{\mrcan;\bm{a}}[\Gamma] = 0.
			\end{equation}
		\end{box_summary}\noindent

	\subsection{Assumption 1: the time-reversal symmetry}		
		\subsubsection{Detailed-balance condition for the intensity functional}
			We refer to this stationary PDF as the {\it canonical distribution} by formally defining the total energy functional $E_{\rm tot}[\Gamma,\bm{a}]$ and the free energy $F(\bm{a})$ based on the following relation: 
			\begin{subequations}
				\label{eq:DB-non-Markovjumps}
				\begin{box_summary}{Assumption 1: Total energy $E_{\rm tot}[\Gamma,\bm{a}]$ and the free energy $F(\bm{a})$}
				\vspace{-5mm}	
				\begin{equation}
					\label{eq:def:equilibrium-dist}
					P_{\mrcan;\bm{a}}[\Gamma] = e^{\beta \left(F(\bm{a})-E_{\rm tot}[\Gamma,\bm{a}]\right)}, \>\>\> 
					e^{-\beta F(\bm{a})} := \int \dd \Gamma e^{-\beta E_{\rm tot}[\Gamma,\bm{a}]}, \>\>\>
					E_{\rm tot}[\Gamma,\bm{a}] = E_{\rm tot}[\Gamma^*,\bm{a}^*],
					\vspace{-1mm}
				\end{equation}
			\end{box_summary}\noindent
			where the energy is assumed to have the even parity. Also, we assume that the initial condition $\hGamma_{t_0}:=(\hx_{t_0},\hH_0):=(\hx_{t_0},\{\hz_{t_0}(s),\hw_{t_0}(s)\}_{s>0})$ is sampled from the canonical distribution $P_{\mrcan;\bm{a}}[\Gamma]$.

			Since we have derived the master equation~\eqref{eq:field_master_hdCP}, we can derive the necessary and sufficient condition for the time-reversal symmetry (see Gardiner's textbook~\cite{GardinerB} and Appendix~\ref{sec:app:detailedbalance}). On the condition that the control parameter $\bm{a}$ is fixed, the necessary and sufficient condition of the time-reversal symmetry is given by\footnote{
			The advective part of the fME operator satisfies the drift condition in Eq.~\eqref{eq:app:detailed-balance-fME} of Appendix~\ref{sec:app:detailedbalance} automatically. Indeed, from Eq.~\eqref{eq:set_complete_dynamics_hdCP}, the advective drifts are $A_{z(s)}[\Gamma]=sw(s)$ and $A_{w(s)}[\Gamma]=-sz(s)$, whose parities are $\eps_{z}=-\eps$ and $\eps_{w}=+\eps$, so that
			\[
				\eps_{z}A_{z(s)}[\Gamma^*] = (-\eps)\,s\,(\eps w(s)) = -A_{z(s)}[\Gamma], \quad
				\eps_{w}A_{w(s)}[\Gamma^*] = \eps\,(-s)\,(-\eps z(s)) = -A_{w(s)}[\Gamma]
			\]
			for both $\eps=\pm 1$. In other words, the advective part of the fME is equal to that for a set of harmonic oscillators and is a time-reversal-invariant Hamiltonian flow. 
			}
			\begin{equation*}
				\Lambda_{\bm{a}}\left[\Gamma \mid \Gamma'\right] P_{\mrcan;\bm{a}}[\Gamma'] = \Lambda_{\bm{a}^*}\left[\Gamma'^{*}\mid\Gamma^{*}\right]P_{\mrcan;\bm{a}}[\Gamma]
				\quad \Longleftrightarrow \quad
				\frac{1}{\beta}\ln \frac{\Lambda_{\bm{a}^*}\left[\Gamma^{*}\mid\Gamma'^{*}\right]}{\Lambda_{\bm{a}}\left[\Gamma' \mid \Gamma\right]} =
				E_{\rm tot}[\Gamma',\bm{a}]-E_{\rm tot}[\Gamma,\bm{a}]
			\end{equation*}
			for the total-system intensity $\Lambda_{\bm{a}}\left[\Gamma'\mid\Gamma \right]$.
			
			This condition can be simplified in the following form: 
			\begin{box_summary}{Assumption 1: Time-reversal symmetry for the target-system intensity}
				\vspace{-0.4cm}
				\begin{equation}
				\begin{aligned}
					\frac{1}{\beta}\ln \frac{\lambda_{\bm{a}^*}\left[\eps(x-x')\mid\Gamma'^{*}\right]}{\lambda_{\bm{a}}\left[x'-x \mid \Gamma\right]} &= 
					E_{\rm tot}[\Gamma',\bm{a}]-E_{\rm tot}[\Gamma,\bm{a}].
					\label{eq:DB-independentOfEmbedding}
				\end{aligned}
				\end{equation}
				\vspace{-0.5cm}	
			\end{box_summary}
		\end{subequations}
		\noindent
		In this paper, the following two relations~\eqref{eq:def:equilibrium-dist} and \eqref{eq:DB-independentOfEmbedding} are called {\it Assumption 1} as the most important form of the time-reversal symmetry.

	\subsection{Assumption 2: the invariance of the total intensity under time reversal}
		According to Gardiner's book~\cite{GardinerB}, Assumptions 0 and 1 are the necessary and sufficient conditions in formulating the time-reversal symmetry. However, the time-reversal symmetry is only a minimal assumption for the consistency with thermodynamics, and it is also conventional to assume the invariance of the total intensity under time reversal:
		\begin{box_summary}{Assumption 2: Invariance of the total intensity under time reversal}
			\vspace{-0.5cm}	
			\begin{equation}
				\lambda_{{\rm tot};\bm{a}}[\Gamma] = \lambda_{{\rm tot};\bm{a}^*}[\Gamma^*], \>\>\> 
				\lambda_{{\rm tot};\bm{a}}[\Gamma] := \int \Lambda_{\bm{a}}[\Gamma'\mid \Gamma]\dd \Gamma' = \int \lambda_{\bm{a}}[y \mid \Gamma]\dd y,
				\vspace{-0.2cm}	
			\end{equation}
		\end{box_summary}
		\noindent
		which we call {\it Assumption 2}. This is a standard assumption that has been made even for Markov stochastic thermodynamics when one studies a system including odd variables.

	\subsection{Solution of the fME under Assumptions 1 and 2}
		By solving the fME~\eqref{eq:field_master_hdCP} under Assumptions 1 and 2, we obtain the general solution in the steady state. This general-solution structure imposes the form of the total energy, such that 
		\begin{box_summary}{Necessary condition as the solution of the fME~\eqref{eq:field_master_hdCP}}
			\vspace{-0.5cm}	
			\begin{gather}\label{eq:gen-sol-fME-canonical}
				E_{\rm tot}[\Gamma,\bm{a}] = \mathcal{G}_{\bm{a}}\left(x, \{r(s),\psi(s)\}_s\right), \>\>\>
				z(s) := r(s)\sin\theta(s), \>\>\>
                w(s) := r(s)\cos\theta(s), \>\>\>
                \psi(s) := \theta(s)-\frac{s}{s^*}\theta(s^*), \\ 
				\label{eq:gen-sol-fME-canonical-PBC}
				\mathcal{G}_{\bm{a}}\left(x, \{r(s),\psi(s)\}_s\right) = 
				\mathcal{G}_{\bm{a}}\left(x, \left\{r(s),\psi(s)+2\pi\left[n(s)-\frac{s}{s^*}n(s^*)\right]\right\}_s\right)
			\vspace{-0.3cm}	
			\end{gather}
		\end{box_summary}\noindent
        with an arbitrary positive value $s^*\in \mathbb{R}^+$ as a reference wave number. This result implies that the energy $E_{\rm tot}[\Gamma,\bm{a}]$ must be the function with the arguments $x$, $\{r(s)\}$, and $\{\psi(s)\}$ as a necessary condition. 

		\subsubsection*{Derivation}
			Under Assumptions 1 and 2, we obtain
			\begin{align}
				\label{eq:sol-fPDE-ME-jump-disappearance}
				\mcL_{\rm J}P_{\mrcan;\bm{a}} &= 
				\int\dd \Gamma' \left\{\Lambda_{\bm{a}}[\Gamma \mid \Gamma']P_{\mrcan;\bm{a}}[\Gamma']-\Lambda_{\bm{a}}[\Gamma' \mid \Gamma]P_{\mrcan;\bm{a}}[\Gamma]\right\} 
				= \int\dd \Gamma' \Lambda_{\bm{a}^*}[\Gamma'^{*} \mid \Gamma^*]P_{\mrcan;\bm{a}}[\Gamma]-\int\dd \Gamma'\Lambda_{\bm{a}}[\Gamma' \mid \Gamma]P_{\mrcan;\bm{a}}[\Gamma] \notag \\
				&= \int\dd \Gamma'^* \Lambda_{\bm{a}^*}[\Gamma'^{*} \mid \Gamma^*]P_{\mrcan;\bm{a}}[\Gamma]-\int\dd \Gamma'\Lambda_{\bm{a}}[\Gamma' \mid \Gamma]P_{\mrcan;\bm{a}}[\Gamma] 
				= \left(\lambda_{{\rm tot};\bm{a}^*}[\Gamma^*]-\lambda_{{\rm tot};\bm{a}}[\Gamma]\right)P_{\mrcan;\bm{a}}[\Gamma] = 0,
			\end{align}
			where we used the invariance under time reversal for the volume element $\dd \Gamma'=\dd \Gamma'^*$ and the total intensity $\lambda_{{\rm tot};\bm{a}}[\Gamma]=\lambda_{{\rm tot};\bm{a}^*}[\Gamma^*]$. Therefore, it is enough to solve the following equation
			\begin{align}
				\mcL_{\rm A}P_{\mrcan;\bm{a}}[\Gamma]  = 
				\int_0^\infty \ds \left\{sz(s)\frac{\delta}{\delta w(s)} -sw(s)\frac{\delta}{\delta z(s)}\right\} P_{\mrcan;\bm{a}}[\Gamma] = 0
                \label{eq:sol-fPDE-ME}
			\end{align}
			in its steady state with the control parameter $\bm{a}$ fixed. 
            
            This equation is a first-order functional differential equation and can be solved by the method of characteristics. Let us consider a variable transformation $\Gamma=(x,\{w(s),z(s)\}_s) \to G_{p}:=(x,\{r(s),\theta(s)\}_s)$:
            \begin{gather}
                z(s)=r(s)\sin\theta(s), \quad w(s)=r(s)\cos\theta(s), \quad 
                r(s):= \sqrt{z^2(s)+w^2(s)}.
            \end{gather}
            Based on this polar coordinate, Eq.~\eqref{eq:sol-fPDE-ME} reduces to 
            \begin{equation}
                \int_0^\infty s\frac{\delta P_{\mrcan;\bm{a}}[G_p]}{\delta \theta(s)}\dd s = 0,\quad 
					P_{\mrcan;\bm{a}}[x,\{r(s),\theta(s)\}] = P_{\mrcan;\bm{a}}[x,\{r(s),\theta(s)+2\pi n(s)\}],
				\end{equation}   
				by imposing the periodic-boundary condition associated with the polar coordinates for any integer-valued function $n(s)$. This periodic boundary condition is the necessary and sufficient condition for the polar-coordinate representation to define a global single-valued PDF. 
            The corresponding Lagrange--Charpit equations and their corresponding solutions are given by 
			\begin{equation}
				\frac{\dd r_\mu(s)}{\dd \mu} = 0,\quad
				\frac{\dd \theta_\mu(s)}{\dd \mu} = -s 
				\quad \Longrightarrow \quad
				r_\mu(s) = A(s), \quad
				\theta_\mu(s) = -s\mu + B(s)
			\end{equation}
            with the parameter $\mu$ for the method of characteristics and constants $A(s), B(s)$ for all $s\in \mathbb{R}^+$. By rewriting 
			\begin{equation}
				C(s):= B(s)-\frac{s}{s^*}B(s^*) = \theta_{\mu}(s) - \frac{s}{s^*}\theta_\mu(s^*)
			\end{equation}
			with a positive reference wave number $s^*>0$. The solution of the Lagrange--Charpit equations implies that the general solution of the original first-order functional differential equation is given by 
			\begin{equation}
				P_{\mrcan;\bm{a}}[G_p] = g_{\bm{a}}(x,\{r(s),\psi(s)\}_s),\quad
                \psi(s):=\theta_{\mu}(s) - \frac{s}{s^*}\theta_\mu(s^*). 
			\end{equation}
			Here the periodic-boundary condition in $\theta(s)$ is equivalent to Eq.~\eqref{eq:gen-sol-fME-canonical-PBC}.

	\subsection{Assumption 3: the exclusive controllability of the target system}
		Let us focus on the class where the controllable part of the energy depends only on the target system $x$, such that
		\begin{box_summary}{Assumption 3: Exclusive controllability of the target system}
			\vspace{-0.5cm}	
			\begin{equation}
				E_{\rm tot}[\Gamma,\bm{a}]:=E_{\rm ctrl}(x,\bm{a}) +  E_{\rm nctrl}[x,\{r(s),\psi(s)\}_s].
				\vspace{-1mm}
			\end{equation}
		\end{box_summary}\noindent
		In the following, we sometimes abbreviate $E_{\rm nctrl}[x,\{r(s),\psi(s)\}_s]$ as $E_{\rm nctrl}$. In the following, this controllability assumption is referred to as {\it Assumption 3}.

	\subsubsection*{Example 1: Zwanzig bath energy}
		The simplest form of the non-controllable energy is given by a quadratic form, 
		\begin{box_summary}{Zwanzig bath energy (ZBE)}
			\vspace{-0.5cm}	
			\begin{equation}
				E_{\rm nctrl}[x_t,\{r_t(s),\psi_t(s)\}_s] := \frac{1}{2}\int_{0}^\infty \dd s\, 
				M(s)r_t^2(s) = \frac{1}{2}\int_{0}^\infty \dd sM(s)\left[w^2_t(s) + z^2_t(s)\right],\label{eq:zbe}
			\end{equation}
			\vspace{-0.6cm}	
		\end{box_summary}\noindent
		which is independent of $\{\psi_t(s)\}_s$ as a special form. Here, $M(s)$ is an arbitrary positive function with a sufficient decay speed (see Sec.~\ref{sec:example:memory-kernel}), physically corresponding to the ``mass'' of the field variables. In this work, this form of non-controllable energy is called the {\it Zwanzig bath energy} (ZBE). The non-Markov jump process constructed from the ZBE is referred to as the {\it Zwanzig jump model} (ZJM); see Sec.~\ref{sec:HowToConstruct} for details. By substituting the solution~\eqref{eq:set_complete_dynamics_initCondition}, this quadratic energy can also be rewritten in the original-variable representation as
		\begin{equation}
		\begin{gathered}
			\label{eq:zbe-original-variable}
			E_{\rm nctrl}[x_t,\{r_t(s),\psi_t(s)\}_s]
			= E_{\rm ini}
			+\int_{t_0}^t \dd \tau\, v_\tau \eta_\tau 
			+\frac{1}{2}\int_{t_0}^t \dd \tau\int_{t_0}^t \dd \tau'\,
			v_\tau \phi(\tau-\tau')v_{\tau'},\\
			E_{\rm ini}:=\frac{1}{2}\int_0^\infty \dd s\,M(s)\left[w_{t_0}^2(s)+z_{t_0}^2(s)\right],\quad
			\eta_\tau:=\int_0^\infty \dd s\,M(s)w_\tau^{\rini}(s),\quad
			\phi(t):=\int_0^\infty \dd s\,M(s)\cos(st).
		\end{gathered}
		\end{equation}
	\subsubsection*{Example 2: Ginzburg--Landau memory-bath energy}
		A more general non-controllable energy can be constructed in the Ginzburg--Landau memory-bath form: 
		\begin{box_summary}{Ginzburg--Landau memory-bath energy}
			\vspace{-0.5cm}	
			\begin{equation}
			\begin{gathered}
				\label{eq:gl-energy}			
				E_{\rm nctrl}[x_t,\{z_t(s),w_t(s)\}_s] := \int_{-\infty}^\infty \dd l \left[\kappa\left(\frac{\pd B_t}{\pd l}\right)^2 + U(B_t)\right], \\
				B_t[l, \{z_t(s),w_t(s)\}_s] = \int_{0}^\infty \dd s M(s)\left[w_t(s)\cos(sl)+z_t(s)\sin(sl)\right].
			\end{gathered}
			\end{equation}
			\vspace{-0.6cm}	
		\end{box_summary}\noindent
		with a non-negative constant $\kappa\geq 0$. The local potential $U$ is chosen such that the canonical PDF $P_{\mrcan;\bm{a}}[\Gamma]$ is normalisable and the energy is invariant under time reversal. See Sec.~\ref{sec:discussion:Rigor} for a more detailed discussion, particularly on its interpretation as the thermodynamic limit of a discrete-mode regularization.

		The memory-bath field $B_t(l)$ can be decomposed into a random term originating from the bath initial condition and a velocity-history term:
		\begin{equation}
		\begin{gathered}
			B_t(l) = \eta_{t-l} + \int_{t_0}^t \phi(t-\tau-l)v_{\tau}\dd \tau, \quad
			\phi(\tau) := \int_{0}^\infty M(s)\cos(s\tau)\dd s, \quad 
			\eta_{t} := \int_0^\infty M(s)w_t^{\rm ini}(s) \dd s.
		\end{gathered}
		\end{equation}
		Note that this model includes the Zwanzig bath energy. Indeed, the quadratic choice $\kappa=0$ and $U(B)=B^2$ gives a Zwanzig-type quadratic bath energy.

\section{Stochastic thermodynamics for non-Markov jump processes}\label{sec:stochasticThermo}
	\subsection{The first law}
	By making Assumptions 1 and 2, let us formulate stochastic thermodynamics for Fourier-embeddable non-Markov jump processes. The first law of thermodynamics is then obtained as an identity:
		\begin{box_summary}{First law of non-Markov stochastic thermodynamics}
			\vspace{-0.5cm}	
			\begin{gather}
				\dd E_{\rm tot}[\hGamma,\bm{a}] = \dd \hW + \dd \hQ, \\
				\dd \hW := \dd \bm{a}\frac{\pd}{\pd \bm{a}}E_{\rm tot}[\hGamma,\bm{a}],\quad 
				\dd \hQ:= \begin{cases}
				\displaystyle E_{\rm tot}[\hGamma',\bm{a}] - E_{\rm tot}[\hGamma,\bm{a}]  & (\mbox{jump from $\Gamma$ to $\Gamma'$: prob.=} \Lambda_{\bm{a}}[\hGamma'\mid \hGamma]\dd\Gamma' \dd t) \\
				0 & (\mbox{no jump: prob.=} 1-\lambda_{{\rm tot};\bm{a}}[\hGamma] \dd t)
				\end{cases}.\notag
			\end{gather}
			\vspace{-5mm}
		\end{box_summary}\noindent

	\subsection{The second law for the total entropy production}
		The second law based on the entropy production is given by 
		\begin{box_summary}{Second law based on the entropy production}
			\vspace{-0.5cm}
			\begin{equation}
				\dsgm_{\rm tot} := \dsgm_{\rm sys} + \dsgm_{\rm bath} \geq 0,
				\vspace{-1mm}
			\end{equation}
		\end{box_summary}\noindent
		where the system and bath entropy productions are defined by 
		\begin{equation}
			\dsgm_{\rm sys} := -\frac{\dd}{\dd t}\int P_t[\Gamma]\ln P_t[\Gamma] \dd \Gamma, \>\>\>
			\dsgm_{\rm bath} := -\beta \left\la \frac{\dd \hat{Q}}{\dd t}\right\ra = \int \Lambda_{\bm{a}}[\Gamma'\mid \Gamma]P_t[\Gamma]\ln \frac{\Lambda_{\bm{a}}[\Gamma' \mid \Gamma]}{\Lambda_{\bm{a}^*}[\Gamma^*\mid \Gamma'^*]}\dd \Gamma \dd \Gamma'.
		\end{equation}
		Note that the total entropy production can be explicitly written as 
		\begin{equation}
			\dsgm_{\rm tot} 
			= \int \dd \Gamma \dd \Gamma' \Lambda_{\bm{a}}[\Gamma'\mid \Gamma]P_t[\Gamma]\ln \frac{\Lambda_{\bm{a}}[\Gamma' \mid \Gamma]P_t[\Gamma]}{\Lambda_{\bm{a}^*}[\Gamma^*\mid \Gamma'^*]P_t[\Gamma']} 
			= \int \dd y\dd \Gamma\lambda_{\bm{a}}[y \mid \Gamma]P_t[\Gamma]\ln \frac{\lambda_{\bm{a}}[y \mid \Gamma]P_t[\Gamma]}{\lambda_{\bm{a}^*}[-\eps y\mid \Gamma'^*]P_t[\Gamma']}.
		\end{equation}
		where $\Gamma'=\Gamma+\Delta\Gamma_y$ in the second expression.

		\subsubsection*{Derivation.} 
			By using the fME, we obtain 
			\begin{equation}
			\begin{aligned}
				\dsgm_{\rm tot} =& -\int \ln P_t[\Gamma]\frac{\pd P_t[\Gamma]}{\pd t}\dd \Gamma +\int \Lambda_{\bm{a}}[\Gamma'\mid \Gamma]P_t[\Gamma]\ln \frac{\Lambda_{\bm{a}}[\Gamma' \mid \Gamma]}{\Lambda_{\bm{a}^*}[\Gamma^*\mid \Gamma'^*]}\dd \Gamma \dd \Gamma' \\
				=& -\int \ln P_t[\Gamma]\left\{sz(s)\frac{\delta}{\delta w(s)}-sw(s)\frac{\delta}{\delta z(s)}\right\}P_t[\Gamma]\dd s\dd \Gamma \\	
				&- \int  \left\{\Lambda_{\bm{a}}[\Gamma\mid \Gamma']P_t[\Gamma']-\Lambda_{\bm{a}}[\Gamma'\mid \Gamma]P_t[\Gamma]\right\}\ln P_t[\Gamma]\dd \Gamma\dd \Gamma' +\int \Lambda_{\bm{a}}[\Gamma'\mid \Gamma]P_t[\Gamma]\ln \frac{\Lambda_{\bm{a}}[\Gamma' \mid \Gamma]}{\Lambda_{\bm{a}^*}[\Gamma^*\mid \Gamma'^*]}\dd \Gamma \dd \Gamma'\\
				=& \int  \Lambda_{\bm{a}}[\Gamma'\mid \Gamma]P_t[\Gamma]\ln \frac{P_t[\Gamma]}{P_t[\Gamma']}\dd \Gamma\dd \Gamma' +\int \Lambda_{\bm{a}}[\Gamma'\mid \Gamma]P_t[\Gamma]\ln \frac{\Lambda_{\bm{a}}[\Gamma' \mid \Gamma]}{\Lambda_{\bm{a}^*}[\Gamma^*\mid \Gamma'^*]}\dd \Gamma \dd \Gamma'\\
				=& \int \dd \Gamma \dd \Gamma'\Lambda_{\bm{a}}[\Gamma'\mid \Gamma]P_t[\Gamma]\ln \frac{\Lambda_{\bm{a}}[\Gamma' \mid \Gamma]P_t[\Gamma]}{\Lambda_{\bm{a}^*}[\Gamma^*\mid \Gamma'^*]P_t[\Gamma']},
			\end{aligned}
			\end{equation}
			where we used 
			\begin{equation}
				\int  \ln P_t[\Gamma]\left\{sz(s)\frac{\delta}{\delta w(s)}-sw(s)\frac{\delta}{\delta z(s)}\right\}P_t[\Gamma]\dd s\dd \Gamma
				= -\int  \left\{s z(s)\frac{\delta}{\delta w(s)} - s w(s)\frac{\delta}{\delta z(s)}\right\}P_t[\Gamma]\dd s\dd \Gamma = 0.
			\end{equation}
			by integration by parts. Finally, by using the inequality $\ln(t^{-1})\geq 1-t$ for all $t>0$ and Assumption 2 $\lambda_{{\rm tot}; \bm{a}}[\Gamma] = \lambda_{{\rm tot}; \bm{a}^*}[\Gamma^*]$,
			we obtain 
			\begin{align}
				\dsgm_{\rm tot} \geq \int \left(1-\frac{\Lambda_{\bm{a}^*}[\Gamma^*\mid \Gamma'^*]P_t[\Gamma']}{\Lambda_{\bm{a}}[\Gamma' \mid \Gamma]P_t[\Gamma]}\right)\Lambda_{\bm{a}}[\Gamma'\mid \Gamma]P_t[\Gamma]\dd \Gamma \dd \Gamma' 
				= \int \lambda_{{\rm tot}; \bm{a}}[\Gamma]P_t[\Gamma]\dd \Gamma - \int \lambda_{{\rm tot}; \bm{a}^*}[\Gamma^*]P_t[\Gamma]\dd \Gamma = 0.
			\end{align}

\subsection{The second law for the irreversible work and its gauge invariance}
	By adding Assumption 3, let us study Helmholtz's free energy. Helmholtz's free energy for this setup is given by 
	\begin{box_summary}{Helmholtz's free energy}
		\vspace{-.5cm}
		\begin{equation}
			F(\bm{a}) := -\beta^{-1}\ln \mc{Z}(\bm{a}), \>\>\> \mc{Z}(\bm{a}) := \int \exp\left(-\beta E_{\rm ctrl}(x,\bm{a})-\beta E_{\rm nctrl}[x,\{z(s),w(s)\}_s]\right)\dd \Gamma.
			\vspace{-1mm}
		\end{equation}		
	\end{box_summary}\noindent
	In the quasi-static limit, we have the identity between the work and the free energy, such that
	\begin{equation}
		\la\Delta \hW\ra_{\rm qs} := \int_{\bm{a}_{\rm ini}}^{\bm{a}_{\rm fin}} \dd \bm{a}\left\la \frac{\pd E_{\rm ctrl}(\hx,\bm{a})}{\pd \bm{a}}\right\ra_{\rm can} = F(\bm{a}_{\rm fin}) - F(\bm{a}_{\rm ini}), \>\>\> 
			\la \hat{A}\ra_{{\rm can}; \bm{a}} := \int A[\Gamma]P_{{\rm can};\bm{a}}[\Gamma]\dd \Gamma.
		\end{equation}
		Here, $\la\Delta\hW\ra_{\rm qs}$ denotes the ensemble-averaged work in the quasi-static limit, where the state is canonical at each $\bm{a}$.
			In addition, let us prove the second law for equilibrium-to-equilibrium transition processes, such that 
	\begin{box_summary}{Second law based on the irreversible work}
		\vspace{-.4cm}
		\begin{equation}
			\sigma_{\rm tot} = \beta \left(\la \Delta \hW \ra -\Delta F\right) \geq 0,\>\>\> \Delta \hW := \int_{\bm{a}_{\rm ini}}^{\bm{a}_{\rm fin}} \dd \bm{a}\frac{\pd E_{\rm ctrl}(\hx,\bm{a})}{\pd \bm{a}},
			\vspace{-.2cm}
		\end{equation}
	\end{box_summary}\noindent
	where we assume that the initial and final states are in thermal equilibrium. Its explicit derivation is given below. 
	
	This equality is important because the cumulative entropy production $\sigma_{\rm tot}$ can be expressed only in terms of the target-system statistics. The second law is therefore gauge invariant under the choice of equilibrium Markov embedding, provided that the embeddings satisfy Assumption 3 and reproduce the same driven target dynamics and marginal equilibrium PDF. A direct proof is given below.

	\subsubsection*{Derivation.} 
		By integrating the second law for the total entropy production for $[t_{\rm i}, t_{\rm f}]$, we obtain 
		\begin{align}
			\sigma_{\rm tot} := \int_{t_{\rm i}}^{t_{\rm f}} \dsgm_{\rm tot}(t)\dd t
			= \sigma_{\rm sys}(t_{\rm f}) - \sigma_{\rm sys}(t_{\rm i}) - \beta \la \Delta \hQ\ra \geq 0.
		\end{align}
		Here we make the assumption that the initial and final states are characterised by the canonical distribution: 
		\begin{equation}
			P_{t_{\rm i}}[\Gamma] = e^{\beta (F(\bm{a}_{\rm i})-E_{\rm tot}[\Gamma,\bm{a}_{\rm i}])}, \>\>\>
			P_{t_{\rm f}}[\Gamma] = e^{\beta (F(\bm{a}_{\rm f})-E_{\rm tot}[\Gamma,\bm{a}_{\rm f}])}.
		\end{equation}
		Under this assumption, the Shannon entropy is given by 
		\begin{equation}
			\sigma_{\rm sys}(t_{\rm f}) - \sigma_{\rm sys}(t_{\rm i})= \beta \left(\left\la \Delta E_{\rm tot}\right\ra -\Delta F\right).
		\end{equation}
		Using the averaged first law $\la\Delta E_{\rm tot}\ra = \la \Delta \hW \ra + \la \Delta \hQ \ra$, we obtain 
		\begin{equation}
			\sigma_{\rm tot} = \beta \left(\la \Delta \hW \ra -\Delta F\right) \geq 0.
		\end{equation}

	\subsubsection*{Explicit proof of the gauge invariance of the cumulative entropy production $\sigma_{\rm tot}$}
		We use the word ``gauge invariance'' to mean that the cumulative entropy production is unchanged under the choice of equilibrium Markov embedding satisfying Assumption 3 and reproducing the same target-system statistics. 

		Let $\chi$ represent the auxiliary variables for an arbitrary equilibrium Markov embedding, such that $\Gamma=(x,\chi)$. Under Assumption 3, we introduce $E'_{\rm nctrl}(x)$ as
		\begin{equation}
			e^{-\beta E'_{\rm nctrl}(x)}:=\int\dd\chi\,e^{-\beta E_{\rm nctrl}(x,\chi)}.
		\end{equation}
		Since $E_{\rm nctrl}(x,\chi)$ is independent of the control parameter $\bm{a}$, $E'_{\rm nctrl}(x)$ is also independent of $\bm{a}$. By integrating the canonical PDF over $\chi$, we obtain the marginal equilibrium PDF of the target system:
		\begin{equation}
			P_{{\rm can};\bm{a}}(x):=\int\dd\chi\,P_{{\rm can};\bm{a}}(x,\chi)
			=e^{\beta\left[F(\bm{a})-E_{\rm ctrl}(x,\bm{a})-E'_{\rm nctrl}(x)\right]}.
			\label{eq:target-canonical-marginal}
		\end{equation}
		We define the experimentally accessible energy $E_{{\rm exp};\bm{a}}(x)$ and free energy $F_{\rm exp}(\bm{a})$ through the marginal equilibrium PDF $P_{{\rm can};\bm{a}}(x)$ for the target system alone: 
		\begin{equation}
			P_{{\rm can};\bm{a}}(x):=e^{\beta\left[F_{\rm exp}(\bm{a})-E_{{\rm exp};\bm{a}}(x)\right]}.
		\end{equation}
		Comparison with Eq.~\eqref{eq:target-canonical-marginal} gives
		\begin{equation}
			E_{{\rm exp};\bm{a}}(x)=E_{\rm ctrl}(x,\bm{a})+E'_{\rm nctrl}(x)+c(\bm{a}),
			\qquad F_{\rm exp}(\bm{a})=F(\bm{a})+c(\bm{a}),
		\end{equation}
		where $c(\bm{a})$ is an arbitrary $\bm{a}$-dependent energy zero. Thus, with $\Delta c:=c(\bm{a}_{\rm fin})-c(\bm{a}_{\rm ini})$,
		\begin{equation}
				\Delta\hW_{\rm exp}:=\int_{\bm{a}_{\rm ini}}^{\bm{a}_{\rm fin}}\dd\bm{a}\frac{\pd E_{{\rm exp};\bm{a}}(x)}{\pd\bm{a}}=\Delta\hW+\Delta c,\quad
				\Delta F_{\rm exp}:=F_{\rm exp}(\bm{a}_{\rm fin})-F_{\rm exp}(\bm{a}_{\rm ini})=\la\Delta\hW_{\rm exp}\ra_{\rm qs}=\Delta F+\Delta c.
		\end{equation}
		We thus obtain
		\begin{equation}
			\sigma_{\rm tot}=\sigma_{\rm exp}:=\beta\left(\la\Delta\hW_{\rm exp}\ra-\Delta F_{\rm exp}\right)
			=\beta\left(\la\Delta\hW\ra-\Delta F\right)\geq0,
		\end{equation}
		which explicitly shows that the arbitrary energy ``gauge'' $c(\bm{a})$ cancels from the cumulative entropy production.

		Differentiating Eq.~\eqref{eq:target-canonical-marginal} with respect to $\bm{a}$ and using the normalisation of $P_{{\rm can};\bm{a}}(x)$, we obtain
		\begin{equation}
			\frac{\pd F(\bm{a})}{\pd\bm{a}}
			=\int\dd x\,P_{{\rm can};\bm{a}}(x)\frac{\pd E_{\rm ctrl}(x,\bm{a})}{\pd\bm{a}}.
		\end{equation}
		Similarly, by introducing the time-dependent target marginal $P_t(x):=\int\dd\chi\,P_t(x,\chi)$, the averaged work is written as
		\begin{equation}
			\la\Delta\hW\ra
			=\int_{t_{\rm i}}^{t_{\rm f}}\dd t\,\frac{\dd\bm{a}_t}{\dd t}
			\int\dd x\,P_t(x)\frac{\pd E_{\rm ctrl}(x,\bm{a}_t)}{\pd\bm{a}_t}.
		\end{equation}
		More directly, Eq.~\eqref{eq:target-canonical-marginal} gives
		\begin{equation}
			-\frac{\pd}{\pd\bm{a}}\ln P_{{\rm can};\bm{a}}(x)
			=\beta\left[\frac{\pd E_{\rm ctrl}(x,\bm{a})}{\pd\bm{a}}-\frac{\pd F(\bm{a})}{\pd\bm{a}}\right].
		\end{equation}
		Therefore, the cumulative entropy production can be expressed only in terms of the target-system statistics as
		\begin{equation}
			\sigma_{\rm tot}
			=\beta\left(\la\Delta\hW\ra-\Delta F\right) 
				=-\int_{t_{\rm i}}^{t_{\rm f}}\dd t\,\frac{\dd\bm{a}_t}{\dd t}
			\int\dd x\,P_t(x)\frac{\pd}{\pd\bm{a}_t}\ln P_{{\rm can};\bm{a}_t}(x).
			\label{eq:target-only-cumulative-EP}
		\end{equation}
		It proves gauge invariance with respect to the choice of equilibrium Markov embedding. In particular, any two equilibrium Markov embeddings satisfying Assumption 3 and reproducing the same $P_t(x)$ and $P_{{\rm can};\bm{a}}(x)$ yield the same cumulative EP for an equilibrium-to-equilibrium transition.

\newpage

\section{Zwanzig jump model}\label{sec:HowToConstruct}
	A natural question arises: how can we construct intensity functionals satisfying Assumptions 1--3? In this section, we show how to systematically construct an intensity functional satisfying Assumptions 1--3, by assuming the ZBE form~\eqref{eq:zbe}. We refer to this model as the Zwanzig jump model (ZJM), reminiscent of the Zwanzig model for the generalised Langevin equation (GLE).

	\subsection{Definition of the Zwanzig jump model (ZJM)}
		For this energy functional, the following intensity functional satisfies Assumptions 1--3: 
		\begin{subequations}				
			\label{eq:example:thermalintensity3}
			\begin{box_summary}{Zwanzig jump model (ZJM; Fourier representation)}
				\vspace{-0.5cm}	
				\begin{align}
					\lambda_{\bm{a}} [y\mid \Gamma] = \lambda_0 \rho_{\bm{a}}(y\mid x)\exp\left[-\frac{\beta}{2}\left\{E_{\rm ctrl}(x',\bm{a})-E_{\rm ctrl}(x,\bm{a})+\int_0^\infty \dd sM(s)\left(\frac{y^2}{2} + yw(s)\right)\right\}\right], 
				\end{align}
				\vspace{-5mm}
				\begin{equation}
					\rho_{\bm{a}}(y\mid x) = \rho_{\bm{a}^*}(\eps y\mid \eps x),\quad
					\rho_{\bm{a}}(y\mid x) = \rho_{\bm{a}^*}(- \eps y\mid \eps x'),\quad 
					y:=x'-x.
					\vspace{-1mm}
				\end{equation}
			\end{box_summary}\noindent
		\end{subequations}
        $M(s)$ is assumed to be any positive function with a sufficient decay speed (see Sec.~\ref{sec:example:memory-kernel}). The non-controllable energy in this construction is the ZBE, and the resulting non-Markov jump process is referred to as the {\it Zwanzig jump model} (ZJM). For the ZJM, the canonical PDF is given by 
        \begin{box_summary}{}
            \vspace{-0.5cm}	
    		\begin{equation}
    			P_{{\rm can};\bm{a}}[\Gamma] = e^{\beta\left(F(\bm{a})-E_{\rm tot}[\Gamma,\bm{a}]\right)},\quad 
    			E_{\rm tot}[\Gamma,\bm{a}] = E_{\rm ctrl}(x,\bm{a}) + \frac{1}{2}\int_0^\infty \dd s M(s)\left[w^2(s)+z^2(s)\right],
    		\end{equation}
            \vspace{-0.6cm}	
		\end{box_summary}\noindent
        which can be checked by its direct substitution into the fME~\eqref{eq:field_master_hdCP}. Also, the stochastic heat is given by 
		\begin{equation}
			\dd Q:= E_{\rm tot}[\Gamma',\bm{a}]-E_{\rm tot}[\Gamma,\bm{a}] = E_{\rm ctrl}(x',\bm{a})-E_{\rm ctrl}(x,\bm{a})+\int_0^\infty \dd sM(s)\left(\frac{y^2}{2} + yw(s)\right).
		\end{equation}

		\subsubsection{Consistency check with Assumption 1}
			Let us check that this model satisfies Assumption 1. Consider $\Gamma:=(x,\{z(s),w(s)\}_s)$ and $\Gamma':=(x',\{z'(s),w'(s)\}_s)$, where $z'(s)=z(s)$ and $w'(s)=(x'-x)+w(s)$. Also, its time reversal is given by 
			\begin{equation}
				\Gamma'^*:=(\eps x', \{z'^*(s), w'^*(s)\}),\quad 
				z'^*(s)=-\eps z(s), \quad 
				w'^*(s)=\eps (x'-x)+\eps w(s).
			\end{equation}
			Thus, with $y:=x'-x$, we have 
			\begin{align}
				\frac{1}{\beta}\ln \frac{\lambda_{\bm{a}^*}[\eps (x-x')\mid \Gamma'^*]}{\lambda_{\bm{a}}[x'-x\mid \Gamma]} 
				=& -\frac{1}{2}\left[ E_{\rm ctrl}(\eps x,\bm{a}^*)-E_{\rm ctrl}(\eps x',\bm{a}^*)+\int_0^\infty \dd sM(s)\left\{\frac{y^2}{2} - \eps y\left(\eps y+\eps w(s)\right)\right\}\right] \notag \\
				&+ \frac{1}{2}\left[ E_{\rm ctrl}(x',\bm{a})-E_{\rm ctrl}(x,\bm{a})+\int_0^\infty \dd sM(s)\left\{\frac{y^2}{2} +yw(s)\right\}\right] 
				+ \frac{1}{\beta}\ln \frac{\rho_{\bm{a}^*}(- \eps y\mid \eps (x+y))}{\rho_{\bm{a}}(y\mid x)} \notag \\
				=& E_{\rm ctrl}(x',\bm{a})-E_{\rm ctrl}(x,\bm{a}) + \int_0^\infty \dd sM(s)\left(\frac{y^2}{2}+yw(s)\right),
			\end{align}
			where we use $E_{\rm ctrl}(\eps x,\bm{a}^*)=E_{\rm ctrl}(x,\bm{a})$, $E_{\rm ctrl}(\eps x',\bm{a}^*)=E_{\rm ctrl}(x',\bm{a})$, and $\rho_{\bm{a}}(y\mid x) = \rho_{\bm{a}^*}(- \eps y\mid \eps (x+y))$. This result shows that the ZJM indeed satisfies Assumption 1. 
			
		\subsubsection{Consistency check with Assumption 2}
			We also check Assumption 2. The total intensity is defined by 
			\begin{align}
				\frac{\lambda_{{\rm tot};\bm{a}}[\Gamma]}{\lambda_0} = \int_{-\infty}^\infty \dd y \rho_{\bm{a}}(y\mid x)\exp\left[-\frac{\beta}{2}\left\{E_{\rm ctrl}(x+y,\bm{a})-E_{\rm ctrl}(x,\bm{a})+\int_0^\infty \dd sM(s)\left(\frac{y^2}{2} + yw(s)\right)\right\}\right]. 
			\end{align}
			For $\Gamma^*:=(\eps x,-\eps z, \eps w)$, the time-reversal total intensity is given by 
			\begin{align}
				\frac{\lambda_{{\rm tot};\bm{a}^*}[\Gamma^*]}{\lambda_0} &= \int_{-\infty}^\infty \dd y \rho_{\bm{a}^*}(y\mid \eps x)\exp\left[-\frac{\beta}{2}\left\{E_{\rm ctrl}(\eps x+y,\bm{a}^*)-E_{\rm ctrl}(\eps x,\bm{a}^*)+\int_0^\infty \dd sM(s)\left(\frac{y^2}{2} + \eps y w(s)\right)\right\}\right]\notag \\
				&= \int_{-\infty}^\infty \dd y \rho_{\bm{a}}(\eps y\mid x)\exp\left[-\frac{\beta}{2}\left\{E_{\rm ctrl}(x+\eps y,\bm{a})-E_{\rm ctrl}(x,\bm{a})+\int_0^\infty \dd sM(s)\left(\frac{y^2}{2} + \eps y w(s)\right)\right\}\right]\notag \\
				&= \int_{-\infty}^\infty \dd y' \rho_{\bm{a}}(y'\mid x)\exp\left[-\frac{\beta}{2}\left\{E_{\rm ctrl}(x+y',\bm{a})-E_{\rm ctrl}(x,\bm{a})+\int_0^\infty \dd sM(s)\left(\frac{y'^2}{2} + y' w(s)\right)\right\}\right],
			\end{align}
			\normalsize
			where we use $\rho_{\bm{a}}(y\mid x) = \rho_{\bm{a}^*}(\eps y\mid \eps x)$ and $E_{\rm ctrl}(\eps x;\bm{a}^*)=E_{\rm ctrl}(x;\bm{a})$ to obtain the second line and apply the variable transformation $y':=\eps y$ to obtain the third line. Thus, Assumption~2 $\lambda_{{\rm tot};\bm{a}}[\Gamma]=\lambda_{{\rm tot};\bm{a}^*}[\Gamma^*]$ is shown.

		\subsubsection{Consistency check with Assumption 0}
			Since the ZJM satisfies Assumptions 1 and 2, the fME admits a canonical-type stationary distribution,
			\begin{equation}
				P_{\mrcan;\bm a}[\Gamma] \propto \exp\left[-\beta E_{\rm ctrl}(x,\bm{a}) - \beta \frac{1}{2}\int_0^\infty \dd s M(s)\left\{z^2(s)+w^2(s)\right\}\right]
			\end{equation}
			as its steady-state solution. The intensity in Eq.~\eqref{eq:example:thermalintensity3} depends on the history only through $\Gamma$ and the stationary solution above satisfies Assumption 0. Thus, the ZJM is Fourier-embeddable and satisfies Assumption~0.
			
			This canonical PDF implies that $x$, $\{z(s),w(s)\}_s$ are independent of each other. Also, it implies that 
            \begin{equation}
                P_{\mrcan;\bm{a}}(x) = \frac{e^{-\beta E_{\rm ctrl}(x,\bm{a})}}{Z_{\rm ctrl}(\bm{a})}, \quad Z_{\rm ctrl}(\bm{a}):= \int_{-\infty}^\infty e^{-\beta E_{\rm ctrl}(x,\bm{a})}\dd x
            \end{equation}
            and $\{z(s),w(s)\}_{s}$ are Gaussian fluctuations that are independent for different modes $s\neq s'$ in the equilibrium state:
            \begin{equation}
                \la \hz(s)\ra_{\rm can} = \la \hw(s)\ra_{\rm can}=0, \quad
                \la \hz(s)\hz(s')\ra_{\rm can}=\la \hw(s)\hw(s')\ra_{\rm can} = \beta^{-1} M^{-1}(s)\delta(s-s'), \quad
                \la \hz(s)\hw(s')\ra_{\rm can} = 0.
				\label{eq:ZJM-canonical-correlations}
            \end{equation}

	\subsection{Intensity functional represented by the original variables $X_t:=\{x_{\tau}\}_{t_0\leq \tau \leq t}$}
		For the ZJM, the intensity can be intuitively rewritten as a functional of $(\hx_t,\{\hv_t\})$: 
		\begin{box_summary}{Zwanzig jump model (ZJM; original variable representation)}
			\vspace{-0.5cm}	
			\begin{gather}
				\label{eq:example:thermalintensity3-xv-rep}
				\lambda_{\bm{a}} [y\mid x, \{v_{\tau}\}_{t_0\leq \tau\leq t},\heta_t] \!=\! \lambda_0 \rho_{\bm{a}}(y\mid x)\exp\left[-\frac{\beta}{2}\!\left\{E_{\rm ctrl}(x+y,\bm{a})\!-\!E_{\rm ctrl}(x,\bm{a})\!+\!\frac{y^2}{2}\phi(0) \!+\! y\int_{t_0}^t  \phi(t-\tau)v_{\tau}\dd \tau +y\heta_t\right\}\!\right], \notag \\
				\phi(t) := \int_0^\infty M(s)\cos (st)\dd s,\quad 
				\heta_t := \int_{0}^\infty M(s)w_t^{\rini}(s)\dd s.
			\end{gather}
		\vspace{-0.5cm}
		\end{box_summary}\noindent
		Here, $\phi(t)$ is the memory kernel and $\heta_t$ is a coloured Gaussian noise term originating from the bath initial condition $H_0=\{\hat z_{t_0}(s),\hat w_{t_0}(s)\}_{s>0}$. The initial state $(\hx_{t_0},H_0)$ is sampled from the canonical PDF~\eqref{eq:def:equilibrium-dist}. The stochastic heat is given by 
		\begin{equation}
			\dd Q := E_{\rm ctrl}(x',\bm{a})-E_{\rm ctrl}(x,\bm{a})+\frac{y^2}{2}\phi(0) + y\int_{t_0}^{t} \phi(t-\tau)v_{\tau}\dd \tau+y\heta_t.
		\end{equation}
		Interestingly, the noise term $\heta_t$ is a coloured Gaussian noise satisfying the fluctuation-dissipation relation (FDR) of the second kind: 
		\begin{box_summary}{Second-kind fluctuation-dissipation relation (FDR) for the Zwanzig jump model}
				\vspace{-0.5cm}	
		\begin{gather}
						\label{eq:FDR-ZJM}
						\la\heta_t\heta_{t+\tau}\ra_{\rm can} = \beta^{-1}\phi(\tau)
				\end{gather}
				\vspace{-0.7cm}
		\end{box_summary}\noindent
		This FDR structure is parallel to that for the GLE. For even parity $\eps=+1$, the ZJM can be interpreted as a jump-process counterpart of the generalized Langevin equation described by Zwanzig's model as shown below.

	\subsection{Microscopic interpretation of the Zwanzig jump model}
		For even parity $\eps=+1$, the ZJM~\eqref{eq:example:thermalintensity3} can be interpreted from the viewpoint of microscopic dynamics. Indeed, let us rewrite the original variables $\{\hz_t(s),\hw_t(s)\}_s$ into the new variables $\{\hq_t(s),\hp_t(s)\}_s$ defined by
		\begin{equation}
			\hq_t(s):= \hx_t-\hw_t(s), \quad \hp_t(s) := \frac{M(s)}{s}\hz_t(s).
		\end{equation}
		The dynamical equations for the new representation $\{\hq_t(s),\hp_t(s)\}_{s}$ are given by
		\begin{equation}
				m(s)\frac{\dd \hq_t(s)}{\dd t} = \hp_t(s), \quad
				\frac{\dd \hp_t(s)}{\dd t} = -s^2 m(s)\left(\hq_t(s)-\hx_t\right),\quad 
				m(s) := \frac{M(s)}{s^2}.
		\end{equation}
		These equations of motion are equivalent to a set of infinitely many harmonic oscillators whose centres are given by the target-state variable $\hx_t$. Indeed, the total energy functional $E_{{\rm tot};\bm{a}}[\hat{G}_t]$ is given by
		\begin{equation}
			E_{{\rm tot};\bm{a}}[\hat{G}_t]
			=E_{\rm ctrl}(\hx_t,\bm{a})
			+\int_0^\infty \dd s
			\left[
				\frac{\hp_t^2(s)}{2m(s)}
				+\frac{m(s)}{2}s^2\left(\hq_t(s)-\hx_t\right)^2
			\right],
			\quad
			\hat{G}_t:=(\hx_t,\{\hq_t(s),\hp_t(s)\}_{s}).
		\end{equation}
		Thus, for $\eps=+1$, the ZJM~\eqref{eq:example:thermalintensity3} can be interpreted as a jump-process counterpart of the Zwanzig model for the GLE, and has a physically sound microscopic foundation.

	\subsection{Examples of the memory kernel}\label{sec:example:memory-kernel}
		We show two concrete forms of $M(s)$, where $M(s)$ is a positive function with a sufficient decay speed. In this work, we assume that it satisfies the following conditions\footnote{If $M(s)=0$ for some modes, those modes are excluded from the phase space.}: 
		\begin{equation}
				M(s) > 0 \quad \mbox{for any }s>0,\quad \lim_{s\to \infty}M(s) = 0,\quad 
				\int_0^\infty M(s)\dd s < \infty.
		\end{equation}
		Note that $\phi(\tau)$ can be represented as a superposition of cosine modes with nonnegative weights if and only if it is positive definite, according to Bochner's theorem~\cite{Franosch2026}.
		
		\paragraph{Gaussian memory kernel.}        
			The following $M(s)$ leads to the Gaussian memory kernel: 
			\begin{equation}
				M_{\rm G}(s):=\frac{2n}{\pi}\exp\left(-\frac{\tau^{*2}s^2}{2}\right)
				\quad \Longrightarrow \quad 
				\phi_{\rm G}(\tau)=\frac{n}{\tau^*}\sqrt{\frac{2}{\pi}}\exp\left(-\frac{\tau^2}{2\tau^{*2}}\right)
			\end{equation}
			with two positive constants, $n:=\int_{0}^\infty \phi(t)\dd t > 0$ and $\tau^*>0$. 
			\vspace{5mm}
		
		\paragraph{Exponential memory kernel.}    
			Also, the following $M(s)$ leads to the exponential memory kernel: 
			\begin{equation}
				M_{\rm exp}(s):=\frac{2}{\pi}\frac{n}{ 1+\tau^{*2} s^2}
				\quad \Longrightarrow \quad 
				\phi_{\rm exp}(\tau)=\frac{n}{\tau^*}e^{-|\tau|/\tau^*}
				\label{eq:example:expon-memory}
			\end{equation}
    		with two positive constants, $n:=\int_{0}^\infty \phi(t)\dd t > 0$ and $\tau^*>0$. This case is very interesting because the ZJM can be efficiently numerically simulated by Laplace embedding, as shown in Sec.~\ref{sec:examples}.
			\vspace{5mm}

		\paragraph{Note on the decay speed of $M(s)$}   
				Since the memory kernel $\phi(\tau)$ is given by $\phi(\tau)=\int_0^\infty M(s)\cos (s\tau)\dd s$, the weight function $M(s)$ must converge to zero at a sufficiently fast speed. Indeed, if the integral of $M(s)$ diverges, the memory kernel $\phi(\tau)$ has a trivial singularity at $\tau=0$: 
				\begin{equation}
						\int_0^\infty M(s)\dd s = \infty \quad \Longrightarrow \quad \phi(0) = \int_0^\infty M(s)\dd s = \infty. 
				\end{equation}
		\newpage

\section{Examples}\label{sec:examples}
	We have considered a general thermodynamic formalism of the non-Markov jump model. Here we consider two concrete examples within the ZJM framework. Note that the models presented in this section cannot be described within the semi-Markov formulation because their intensity functionals depend on their entire histories. Our numerical simulations are based on the thinning algorithm~\cite{Ross2023}---a standard method for efficiently simulating non-homogeneous Poisson processes.

	\subsection{Example 1: Two-state Zwanzig jump model}
		\begin{figure*}
			\centering
			\includegraphics[width=120mm]{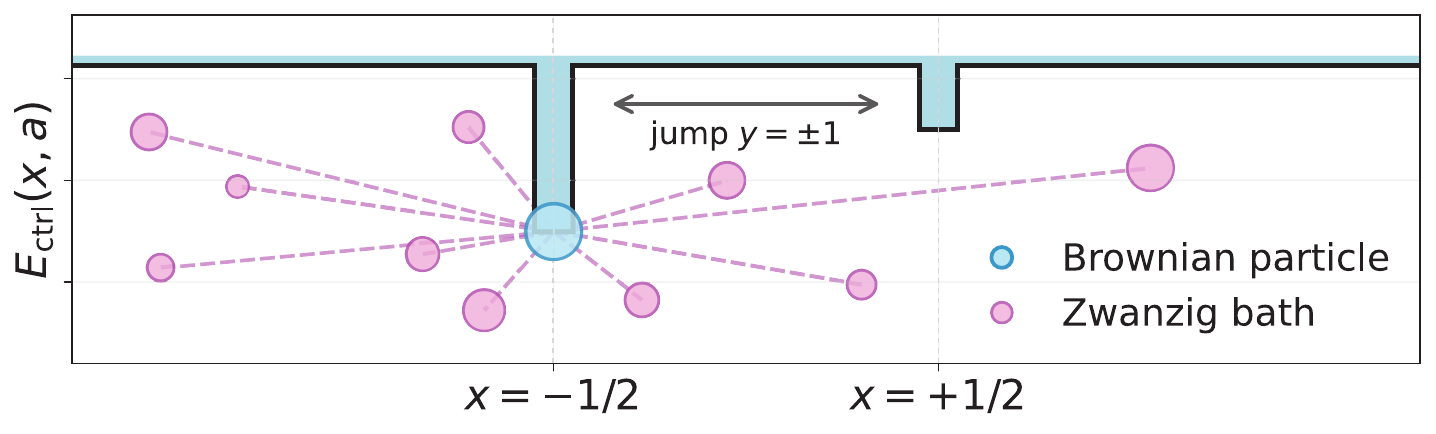}
			\caption{
				Microscopic interpretation of the two-state Zwanzig jump model~\eqref{eq:example:two-state-Zwanzig} for even parity $\eps=+1$. A Brownian particle (blue ball) is trapped by the double-well potential at $x=\pm 1/2$ and is surrounded by the Zwanzig-bath particles (pink balls). The interaction between the Brownian particle and the Zwanzig-bath particles is modelled by harmonic interactions. Thus, this jump model can be regarded as a two-state jump version of the Zwanzig model, originally for the GLE.  
			}
			\label{fig:two-state-Zwanzig-model}
		\end{figure*}
	    	Suppose a two-state system $\hx_t=\pm 1/2$ with parity $\hx_t^*=\eps\hx_t$, where $\eps=\pm 1$, modelled by ZJM dynamics. We assume that the external parameter has the same parity, $a_t^*=\eps a_t$, so that the controllable energy is invariant under time reversal. Let us consider the following energy functional:
		\begin{box_summary}{Energy functional of the two-state Zwanzig jump model}
			\vspace{-0.5cm}	
			\begin{equation}
				\hx_t = \pm \frac{1}{2}, \quad
				E_{\rm ctrl}(\hx_t,a_t) = -a_t\hx_t, \quad
				E_{\rm nctrl}[\hz_t,\hw_t] = \frac{1}{2}\int_0^\infty M(s)\left[\hw_t^2(s)+\hz_t^2(s)\right]\dd s.
			\end{equation}
			\vspace{-0.6cm}	
		\end{box_summary}\noindent
		For even parity $\eps=+1$, the corresponding Brownian-particle setup is illustrated in Fig.~\ref{fig:two-state-Zwanzig-model}, with an even control parameter $a_t^*=a_t$. For odd parity $\eps=-1$, the same model instead represents a classical spin in a magnetic field $a_t$, with $\hx_t^*=-\hx_t$ and $a_t^*=-a_t$.
		We refer to this model as the {\it two-state Zwanzig jump model}. Since the jump size $\hy$ must be $-1$ ($+1$) when $\hx_t=+1/2$ ($\hx_t=-1/2$), the transition rate has the following intensity form:
		\begin{equation}
		\begin{gathered}
			\lambda_{\bm{a}} [y \mid \Gamma] = \rho(y\mid x)\lambda_{\bm{a};y}[\Gamma], \quad 
				\lambda_{\bm{a};y}[\Gamma] = \tilde{\lambda}_0 \exp\left[-\frac{y \beta}{2}
				\left\{
					-a
					+\int_0^\infty  M(s) w(s)\dd s
				\right\}\right], \\
			\rho(y\mid x):= \mathbb{I}_{x=-1/2}\delta(y-1) + \mathbb{I}_{x=+1/2}\delta(y+1),\quad 
			\tilde{\lambda}_0:=\lambda_0\exp\left[-\frac{\beta}{4}\int_0^\infty M(s)\dd s \right],
		\end{gathered}
		\end{equation}
		where we used relations $w'(s):=y+w(s)$ and $y^2=1$. This intensity functional satisfies Assumptions 1--3. This intensity is equivalent to the following original-variable representation:	
		\begin{box_summary}{Intensity functional for the two-state Zwanzig jump model}
			\vspace{-0.5cm}	
			\begin{equation}
			\begin{gathered}
				\label{eq:example:two-state-Zwanzig}
				\lambda_{\bm{a}}[y \mid \hx, \{\hv_\tau\}_{t_0\leq \tau\leq t}, \heta_t] = \tilde{\lambda}_0 \rho(y\mid \hx)\exp\left[-\frac{y \beta}{2}
				\left\{
					-a
					+\int_{t_0}^{t}  \phi(t-\tau)\hv_{\tau}\dd \tau + \heta_t
				\right\}\right],\\ 
				\phi(\tau):=\int_0^\infty M(s)\cos(s\tau)\dd s, \quad
				\heta_t := \int_0^\infty M(s)\hw_t^{\rini}(s)\dd s,\quad 
				\la \heta_t\heta_{t+\tau}\ra_{\rm can} = \beta^{-1}\phi(\tau).
			\end{gathered}
			\end{equation}
			\vspace{-0.6cm}	
		\end{box_summary}

		The two-state Zwanzig jump model is closely related to the generalised Glauber dynamics of Ref.~\cite{ChenPRX2023}. It was introduced for biological inference and applied to the Eco-HAB experiment, where the history-dependent chamber-occupancy dynamics of 15 interacting mice in a four-chamber habitat was modelled by representing each mouse as a four-state Potts variable.
		The generalised Glauber dynamics is Fourier-embeddable because its harmonic-oscillator bath generates a discrete cosine-mode representation of the memory. With its canonical bath preparation and reversible transition rates, it satisfies Assumptions 0--2 under the even-variable convention, and Assumption 3 for target-only energetic control.

		The stochastic work and heat are given by 
		\begin{equation}
			\dd \hW := -\frac{\dd a_t}{\dd t}\hx_t \dd t, \quad 
			\dd \hQ := -a_t\hy  + \hy\int_{t_0}^{t} \phi(t-\tau)\hv_{\tau}\dd \tau + \frac{1}{2}\phi(0) + \hy\heta_t
		\end{equation}
		for a jump size $\hy$. The second law is given by 
		\begin{equation}
			\sigma_{\rm tot}=\beta \left(\la \Delta \hW\ra-\Delta F\right) \geq 0,\quad 
			\Delta F = F(a_{\rm fin}) - F(a_{\rm ini}) =  -\beta^{-1}\ln \frac{\cosh(\beta a_{\rm fin}/2)}{\cosh(\beta a_{\rm ini}/2)}.
			\label{ineq:2ndLaw:two-level-example}
		\end{equation}
				\begin{figure*}
			\centering 
			\includegraphics[width=180mm]{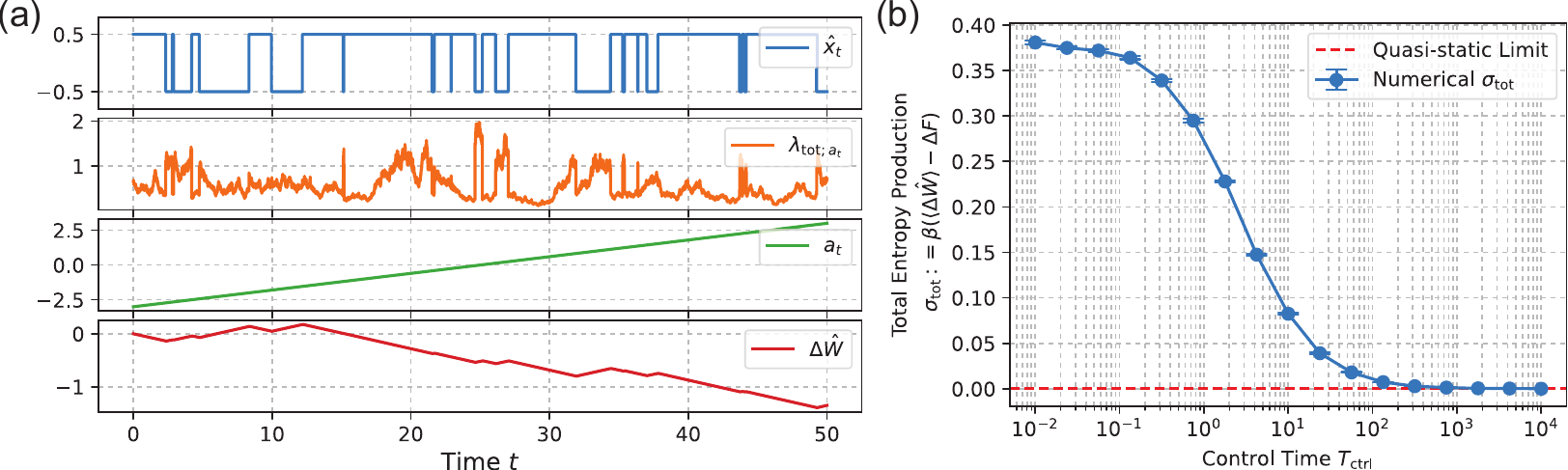}
			\caption{
				Numerical verification of the second law $\sigma_{\rm tot}\geq 0$ for the two-state Zwanzig jump model with $a_{\rm ini}=-3.0$, $a_{\rm fin}=3.0$, $\tilde{\lambda}_0=0.5$, $\beta=0.3$, $n=6$, and $\tau^*=2$.  
				(a)~Sample paths for the target system $\hx_t$, the intensity $\hat{\lambda}(t)$, the external parameter $a_t$, and the stochastic work $\Delta \hat{W}$. The protocol is linear $a_t=a_{\rm ini} + t(a_{\rm fin}-a_{\rm ini})/T_{\rm ctrl}$ with the control time $T_{\rm ctrl}=50$.
				(b)~The total entropy production $\sigma_{\rm tot}:=\beta (\la\Delta \hat{W}\ra-\Delta F)$ between two equilibrium states characterised by $a_{\rm ini}$ and $a_{\rm fin}$. The total entropy production $\sigma_{\rm tot}$ satisfies the second law $\sigma_{\rm tot}\geq 0$ as predicted by our formulation. The number of Monte Carlo iterations is $N_{\rm MC}=2\times 10^5$. Error bars denote the standard error of the mean.
			}
			\label{fig:TwoLevel-Numerical}
		\end{figure*}

		\subsubsection*{Numerical simulation}
			Let us consider a numerical simulation for the two-state Zwanzig jump model, particularly by focusing on the exponential memory case~\eqref{eq:example:expon-memory}. Since the coloured Gaussian noise $\heta_t$ appears in the original-variable representation, the numerical simulation can be efficiently performed by using the Laplace embedding
			\begin{equation}
				\hZ_t := \frac{n}{\tau^*}\int_{-\infty}^{t}e^{-(t-\tau)/\tau^*}\hv_{\tau}\dd \tau + \heta_t, \quad \la \heta_{t} \heta_{t+\tau}\ra = \frac{n}{\beta \tau^*}e^{-|\tau|/\tau^*}
			\end{equation}
			where the initial condition is set to be in the equilibrium state, $t_0=-\infty$. The variable $\hZ_t$ obeys the following closed stochastic equation:
			\begin{equation}
				\frac{\dd \hZ_t}{\dd t} = -\frac{\hZ_t}{\tau^*} + \frac{n}{\tau^*}\hxi^{\rm CP}_{\lambda(y\mid \hat{G}_t)} + \sqrt{\frac{2n}{\beta\tau^{*2}}}\hxi^{\rm G},\quad
				\hat{G}_t:=(\hx_t,\hZ_t),
			\end{equation}
			where $\hxi_t^{\rm G}$ is the Gaussian white noise independent of the jump noise and satisfies $\la\hxi_t^{\rm G}\hxi_{t'}^{\rm G}\ra=\delta(t-t')$. The intensity is simplified as 
			\begin{equation}
				\lambda_{\bm{a}}(y \mid G_t):= \rho(y \mid x)\lambda_{a;y}(Z_t), \quad
				\lambda_{a;y}(Z_t):= \tilde{\lambda}_0 \exp \left\{-\frac{y\beta}{2}\left(-a + Z_t\right)\right\}.
			\end{equation}
			Thus, by selecting the stochastic Laplace embedding, this non-Markov system can be formulated as the two-dimensional Markov process, whose master equation is given by 
			\begin{equation}
				\begin{split}
				\frac{\pd P_t(x,Z)}{\pd t} =& \frac{1}{\tau^*}\frac{\pd}{\pd Z} \left[Z P_t(x,Z)\right]
				+\frac{n}{\beta\tau^{*2}}\frac{\pd^2 P_t(x,Z)}{\pd Z^2} \\
				&+\sum_{y=\pm 1}\left[\lambda_{a;y}(Z-yn/\tau^*)\mathbb{I}_{x=y/2}P_{t}(x-y,Z-yn/\tau^*) - \lambda_{a;y}(Z)\mathbb{I}_{x=-y/2}P_{t}(x,Z)\right].
				\end{split}
			\end{equation}
			Thus, for the exponential memory kernel, the Laplace embedding provides a compact simulation scheme for the same ZJM dynamics. See a sample path of $\lambda_{a_t}$ in Fig.~\ref{fig:TwoLevel-Numerical}(a).
			
			The thermodynamic quantities are equivalently evaluated in this compact representation, because $\hZ_t$ represents the sum of the memory force and the equilibrium Gaussian noise. The stochastic work and heat are given by 
				\begin{equation}
					\dd \hW := -\frac{\dd a_t}{\dd t}\hx_t\dd t, \quad 
					\dd \hQ := \hy(-a_t + \hZ_t) + \frac{n}{2\tau^*}
				\end{equation}
			for the jump size $\hy$. The second law is given by 
			\begin{equation}
				\sigma_{\rm tot}=\beta \left(\la \Delta \hat{W}\ra-\Delta F\right) \geq 0.
				\label{ineq:numerical:2ndlaw-two-level}
			\end{equation}
			This fact demonstrates that the numerical implementation can be simplified without changing the thermodynamic observables. 

			We performed the Monte Carlo numerical simulation of this model based on the representation $\hat{G}_t=(\hx_t, \hZ_t)$. The external parameter $a(t)$ is deterministically controlled according to the following schedule: 
			\begin{equation}\label{eq:protocol-linear}
				a_t = \begin{cases}
					a_{\rm ini} & (T_{\rm start}\leq t\leq  0) \\
					\displaystyle \frac{t}{T_{\rm ctrl}}(a_{\rm fin}-a_{\rm ini}) + a_{\rm ini} & (0<t<T_{\rm ctrl}) 
				\end{cases},
			\end{equation}
			where $T_{\rm start}$ is the equilibration time of the numerical simulation and 
			$T_{\rm ctrl}$ is the total time of the external parameter control. The initial condition is given by $x_{T_{\rm start}}=\pm 1/2$ (with the equilibrium distribution) and $\hZ_{T_{\rm start}}$ sampled from the stationary Gaussian distribution with variance $n/(\beta\tau^*)$. Theoretically, by taking the limit $T_{\rm start}\to -\infty$, the system can be regarded as being in the equilibrium state at $t=0$. We therefore take a very large negative number for $T_{\rm start}$. For the $i$th Monte Carlo iteration, we numerically evaluated the irreversible work 
			\begin{equation}
				\Delta \hW_i:= -\frac{a_{\rm fin}-a_{\rm ini}}{T_{\rm ctrl}}\int_{0}^{T_{\rm ctrl}} \hx_t\dd t.
			\end{equation}
			We evaluated its ensemble average to obtain the total entropy production (or equivalently, the irreversible work) as 
			\begin{equation}\label{eq:numerical-EP}
				\sigma_{\rm tot}=\beta \left(-\Delta F+\frac{1}{N_{\rm MC}}\sum_{i=1}^{N_{\rm MC}}\Delta \hW_i\right), \quad 
				\Delta F := -\beta^{-1}\ln \frac{\cosh(\beta a_{\rm fin}/2)}{\cosh(\beta a_{\rm ini}/2)},
			\end{equation}
			where the total number of the Monte Carlo iterations, represented by $N_{\rm MC}$, is assumed to be sufficiently large. If $T_{\rm ctrl}$ is sufficiently large (i.e., $T_{\rm ctrl}\gg \max \{1/\tilde{\lambda}_0, \tau^*\}$), the control process can be regarded as being quasi-static, and the entropy production should be zero: $\lim_{T_{\rm ctrl}\to \infty}\sigma_{\rm tot} = 0$.
			
\newpage
			
\subsection{Example 2: non-Markov random-walk model under an external potential}
	\begin{figure*}
		\centering
		\includegraphics[width=120mm]{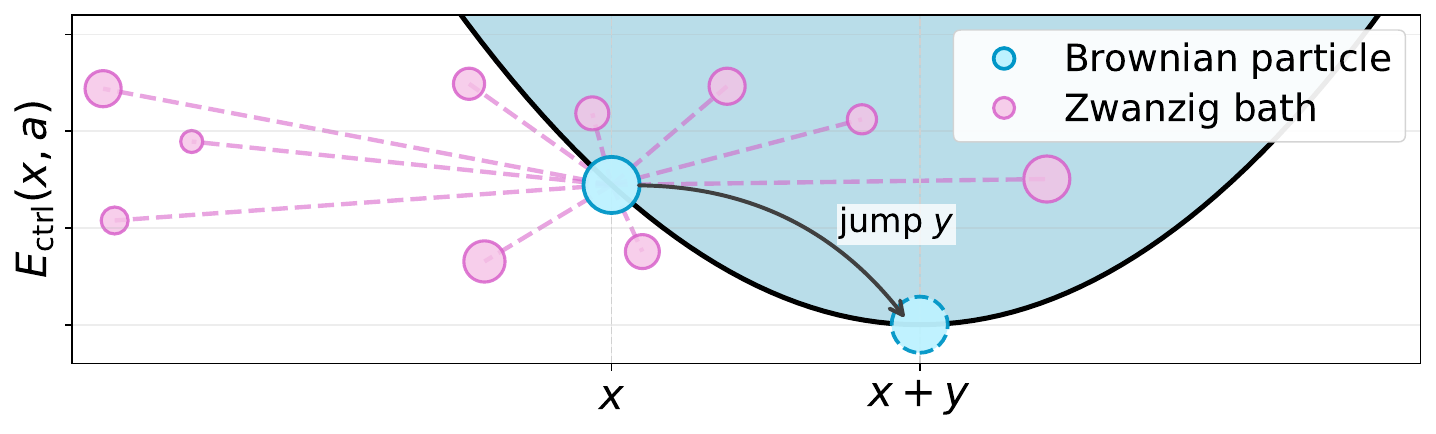}
		\caption{
			Microscopic interpretation of the random-walk Zwanzig jump model~\eqref{eq:def:ZJM-RW}. A Brownian particle exhibiting random-walk jumps (blue ball) is surrounded by the Zwanzig-bath particles (pink balls). The interaction between the Brownian particle and the Zwanzig-bath particles is modelled by harmonic interactions. Thus, this jump-type random-walk model can be regarded as a variant of the Zwanzig model.  
		}
		\label{fig:RW-Zwanzig-model}
	\end{figure*}
	As another example, we consider a non-Markov random-walk model under an external harmonic potential (see Fig.~\ref{fig:RW-Zwanzig-model} for its schematic microscopic setup). Let us consider a position $\hx_t$ with even parity $\eps=1$ within the ZJM, obeying the following energy function: 
	\begin{box_summary}{Energy functional of the ZJM for the non-Markov random-walk model}
		\vspace{-0.5cm}	
		\begin{equation}
			E_{\rm ctrl}(\hx_t,a_t) = \frac{a_t}{2}\hx^2_t, \quad
			E_{\rm nctrl}[\hz_t,\hw_t] = \frac{1}{2}\int_0^\infty M(s)\left[\hw_t^2(s)+\hz_t^2(s)\right]\dd s,
		\end{equation}
		\vspace{-0.5cm}	
	\end{box_summary}\noindent
	where $a_t>0$ and the potential has an even parameter $a^*=a$. Let us consider the following intensity functional, 
	\begin{equation}
		\lambda_{\bm{a}} [y\mid \Gamma] = \lambda_0 \rho(y)\exp\left[-\frac{\beta}{2}\left\{E_{\rm ctrl}(x+y,\bm{a})-E_{\rm ctrl}(x,\bm{a})+\int_0^\infty \dd sM(s)\left(\frac{y^2}{2} + yw(s)\right)\right\}\right]
	\end{equation}
	with $\rho(y)=\rho(-y)$, or equivalently,
		\begin{box_summary}{ZJM intensity functional for the non-Markov random-walk model}
			\vspace{-0.5cm}
			\begin{equation}\label{eq:def:ZJM-RW}
			\begin{gathered}
				\lambda_{a}[y\mid \hx,\{\hv_\tau\}_{t_0\leq \tau\leq t}, \heta_t] = \lambda_0 \rho(y)\exp\left[-\frac{\beta}{2}\left\{E_{\rm ctrl}(\hx+y,a)-E_{\rm ctrl}(\hx,a)+\frac{y^2}{2}\phi(0)+y\int_{t_0}^{t} \phi(t-\tau)\hv_{\tau}\dd \tau+y\heta_t\right\}\right],\\
				\phi(\tau):=\int_0^\infty M(s)\cos(s\tau)\dd s, \quad
				\heta_t := \int_0^\infty M(s)\hw_t^{\rini}(s)\dd s,\quad 
				\la \heta_t\heta_{t+\tau}\ra_{\rm can} = \beta^{-1}\phi(\tau).
				\vspace{-0.6cm}	
			\end{gathered}
			\end{equation}
		\end{box_summary}\noindent
		This model satisfies Assumptions 1--3. The stochastic work and heat are given by 
		\begin{equation}
			\dd \hW := \frac{1}{2}\frac{\dd a_t}{\dd t}\hx_t^2\dd t,\quad 
			\dd \hQ := E_{\rm ctrl}(\hx+\hy,a_t)-E_{\rm ctrl}(\hx,a_t) + \frac{\hy^2}{2}\phi(0) + \hy\int_{t_0}^{t} \phi(t-\tau)\hv_{\tau}\dd \tau+\hy\heta_t.
		\end{equation}
		The second law is given by 
		\begin{equation}
			\sigma_{\rm tot}= \beta\left(\la\Delta \hat{W}\ra-\Delta F\right) \geq 0,\quad 
			\Delta F := F(a_{\rm fin}) - F(a_{\rm ini}) = \frac{\beta^{-1}}{2}\ln \frac{a_{\rm fin}}{a_{\rm ini}}.
		\end{equation}

		\subsubsection*{Numerical simulation}
			\begin{figure*}
				\centering 
				\includegraphics[width=180mm]{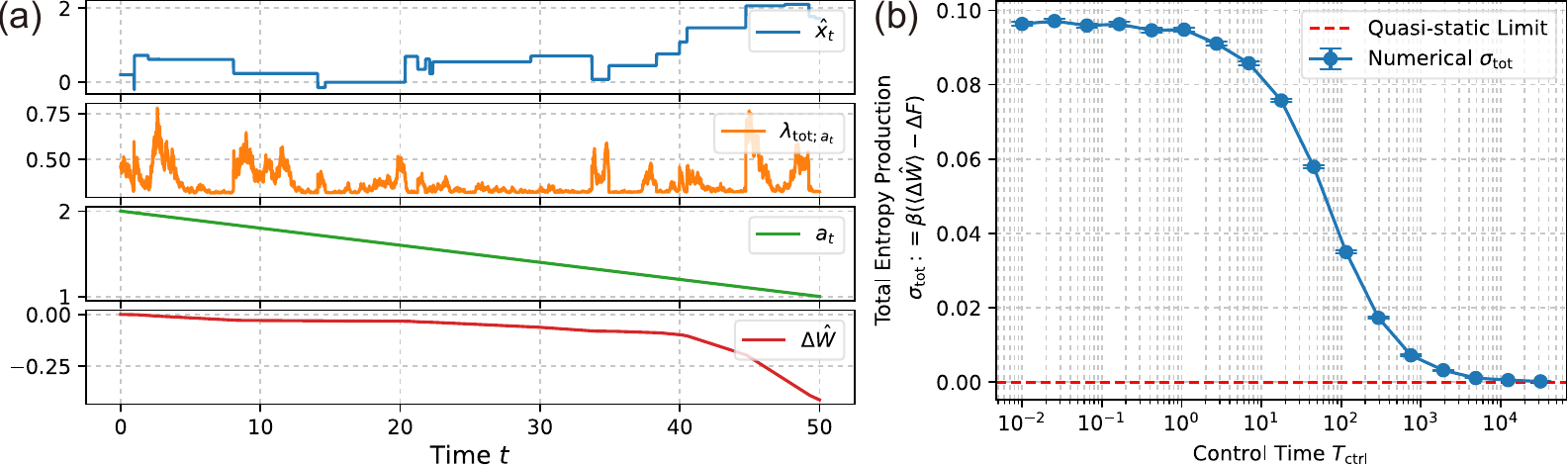}
				\caption{
					Numerical verification of the second law $\sigma_{\rm tot}\geq 0$ for the non-Markov random-walk model with $a_{\rm ini}=2.0$, $a_{\rm fin}=1.0$, $\lambda_0=0.5$, $\beta=1.0$, $n=30$, $\tau^*=3$, and $\sigma=0.5$.  
					(a)~Sample paths for the target system $\hx_t$, the total intensity $\hat{\lambda}_{\rm tot}(t)$, the external parameter $a_t$, and the stochastic work $\Delta \hat{W}$. The protocol is linear $a_t=a_{\rm ini} + t(a_{\rm fin}-a_{\rm ini})/T_{\rm ctrl}$ with the control time $T_{\rm ctrl}=50$.
					(b)~The total entropy production $\sigma_{\rm tot}:=\beta (\la\Delta \hat{W}\ra-\Delta F)$ between two equilibrium states characterised by $a_{\rm ini}$ and $a_{\rm fin}$. The total entropy production $\sigma_{\rm tot}$ satisfies the second law $\sigma_{\rm tot}\geq 0$ as predicted by our formulation. The number of Monte Carlo iterations is $N_{\rm MC}=4\times 10^5$. Error bars denote the standard error of the mean.
				}
				\label{fig:NonMarkovRandomWalks-Numerical}
			\end{figure*}
			Let us perform the numerical simulation of the non-Markov random walk, particularly by focusing on the exponential memory case~\eqref{eq:example:expon-memory} and the Gaussian distribution for $\rho(y)$ (see Fig.~\ref{fig:NonMarkovRandomWalks-Numerical}(a) for its sample paths). By introducing the stochastic Laplace embedding
			\begin{equation}
				\hZ_t := \frac{n}{\tau^*}\int_{t_0}^{t}e^{-(t-\tau)/\tau^*}\hv_{\tau}\dd \tau+\heta_t, \quad 
				\frac{\dd \hZ_t}{\dd t} = -\frac{\hZ_t}{\tau^*} + \frac{n}{\tau^*}\frac{\dd \hx_t}{\dd t} + \sqrt{\frac{2n}{\beta\tau^{*2}}}\hxi_t^{\rm G},
			\end{equation}
			where $\hat{G}_t:=(\hx_t,\hZ_t)$ and $\hxi_t^{\rm G}$ is the Gaussian white noise satisfying $\la\hxi_t^{\rm G}\hxi_{t'}^{\rm G}\ra=\delta(t-t')$. We can perform an efficient numerical simulation based on the stochastic Laplace embedding representation with a simplified intensity
			\begin{equation}
				\lambda_{a_t}(y\mid G_t) = \lambda_0 \rho(y)\exp\left[-\frac{\beta}{2}\left\{\frac{a_t}{2}y^2 + a_t x_t y+\frac{ny^2}{2\tau^*}+yZ_t\right\}\right], 
				\quad 
				\rho(y) = \frac{1}{\sqrt{2\pi \sigma^2}}\exp\left[-\frac{y^2}{2\sigma^2}\right].
			\end{equation}
			The total intensity is given by 
			\begin{equation}
				\lambda_{{\rm tot}; a_t}(G_t) = \int_{-\infty}^\infty \lambda_{a_t}(y\mid G_t)\dd y
				= \frac{\lambda_0}{\sqrt{A_t}}\exp\left[\frac{\beta^2 \sigma^2 (a_tx_t + Z_t)^2}{8A_t}\right], \quad 
				A_t := 1+\frac{\beta \sigma^2 a_t}{2} + \frac{\beta \sigma^2 n}{2\tau^*}
			\end{equation}
			Assuming the occurrence of a jump, its size $y$ obeys the Gaussian distribution 
			\begin{equation}
				P(y\mid G_t, a_t) = \frac{\lambda_{a_t}(y\mid G_t)}{\lambda_{{\rm tot}; a_t}(G_t)}= \frac{1}{\sqrt{2\pi\sigma^2/A_t}}\exp\left[-\frac{A_t}{2\sigma^2}\left\{y+\frac{\beta \sigma^2}{2A_t}(a_t x_t+Z_t)\right\}^2\right].
			\end{equation}
		We then apply the linear protocol~\eqref{eq:protocol-linear} to evaluate the stochastic work 
		\begin{equation}
			\Delta \hW:= \frac{a_{\rm fin}-a_{\rm ini}}{2T_{\rm ctrl}}\int_{0}^{T_{\rm ctrl}} \hx_t^2\dd t.
		\end{equation}
		We obtain the total entropy production using the estimator analogous to Eq.~\eqref{eq:numerical-EP}, with the free-energy difference given above. Finally, we numerically verify the irreversible-work inequality $\sigma_{\rm tot}=\beta(\la\Delta \hat{W}\ra-\Delta F)\geq 0$ as shown in Fig.~\ref{fig:NonMarkovRandomWalks-Numerical}(b).

\newpage

\section{Discussion}\label{sec:discussion}

	\subsection{Remark on the mathematical rigor}\label{sec:discussion:Rigor}
		While our framework is clear enough at the theoretical-physics level, presenting its mathematically rigorous formulation is out of scope of this work. The functional MEs and the functional Fokker--Planck equations have been used in theoretical physics particularly for systems described by the SPDEs~\cite{GardinerB}. However, its rigorous mathematical formulation has been missing\footnote{Particularly, if the SPDEs are nonlinear, the ultraviolet divergence is known to appear. Note that our SPDEs are linear and do not have the renormalisation problems associated with nonlinear SPDEs.}.

		The safe interpretation is that we regard the functional ME as the formal limit from the lattice model~\cite{GardinerB}. This approach is feasible for the non-Markov jump processes and makes sense for our stochastic-thermodynamic formulation. Indeed, we can consider the case where only finite components are important regarding the wave numbers (like the example of the two-state system in Sec.~\ref{sec:examples}): 
		\begin{equation}
			\hlambda = \lambda_{\bm{a}}(y\mid \hGamma_t), \>\>\>
			\hGamma_t := (\hx_t, \hbz_t, \hbw_t), \>\>\>
			\hbz_t := (\hz_{1;t},\dots,\hz_{K;t}), \>\>\> 
			\hbw_t := (\hw_{1;t},\dots,\hw_{K;t}),
		\end{equation}
		where 
		\begin{equation}
			\hz_{k;t} := \hz_{k;t}^{\rm ini}+\int_{t_0}^{t}\hv_{\tau}\sin\{s_k(t-\tau)\}\dd \tau,\quad
			\hw_{k;t} := \hw_{k;t}^{\rm ini}+\int_{t_0}^{t}\hv_{\tau}\cos\{s_k(t-\tau)\}\dd \tau
		\end{equation}
		with a finite positive integer $K>0$. Then, the fME reduces to 
		\begin{align}
			\frac{\pd P_t(\Gamma)}{\pd t} = \sum_{k=1}^K \left[s_k z_k\frac{\pd}{\pd w_k} -s_k w_k\frac{\pd }{\pd z_k}\right]P_t(\Gamma) 
			+ \int_{-\infty}^\infty \dd y \left[\lambda_{\bm{a}}(y\mid \Gamma-\Delta\Gamma_y)P_t(\Gamma-\Delta\Gamma_y)-\lambda_{\bm{a}}(y\mid \Gamma)P_t(\Gamma)\right]
			\label{eq:finiteD-fME}
		\end{align}
		with the difference vector $\Delta \Gamma_y := (y, \bm{0}, y\bm{1})$, $\bm{s}:=(s_1,...,s_K)$, $\bm{0}:=(0,\dots,0)$, and $\bm{1}:=(1,...,1)$. 
        
		This equation is a conventional ME for a finite vector space, and, thus, there is no problem regarding mathematical rigor. In addition, the first and second laws for stochastic thermodynamics hold for an arbitrary $K$. Thus, all the final results remain valid regarding the thermodynamic structure presented in our work for any large $K$. 

        \subsubsection*{Example: non-Markov random-walk model with an approximated exponential memory}
            As a concrete example, let us consider a finite-dimensional version of the non-Markov random-walk model: 
            \begin{subequations}
                \label{eq:finiteD-nonMarkovRW-exp}
                \begin{equation}
                    \lambda_a(y\mid \hGamma_t) = \lambda_0 \rho(y)\exp\left[-\frac{\beta}{2}\left\{E_{\rm ctrl}(x_t+y,a)-E_{\rm ctrl}(x_t,a)+\sum_{k=1}^K M_k\left(\frac{y^2}{2}+y\hw_{k;t}\right)\right\}\right],\quad 
                    \rho(y) = \frac{1}{\sqrt{2\pi \sigma^2}}e^{-y^2/(2\sigma^2)}.
                \end{equation}
				This model can be described by a $(2K+1)$-dimensional master equation and can be numerically simulated as a $(2K+1)$-dimensional SDE. At the same time, this intensity is equivalent to 
                \begin{align}
                    \lambda_a[y\mid \{\hv_{\tau}\}_{t_0\leq \tau\leq t},\heta_t] = \lambda_0 \rho(y)\exp\left[-\frac{\beta}{2}\left\{
                        E_{\rm ctrl}(\hx_t+y,a)-E_{\rm ctrl}(\hx_t,a)+\frac{y^2}{2}\phi(0) + y\int_{t_0}^{t} \phi(t-\tau)\hv_{\tau}\dd \tau+y\heta_t
                    \right\}\right],
                \end{align}
                with the memory kernel 
                \begin{equation}
                    \phi(\tau) = \sum_{k=1}^K M_k \cos(s_k\tau)
                \end{equation}
				and the noise term $\heta_t:=\sum_{k=1}^K M_k\hw_{k;t}^{\rm ini}$. Evidently, any positive-definite memory kernel can be approximated by considering a sufficiently large $K$. For example, an exponential memory can be approximated by 
                \begin{equation}
                    M_k = \frac{2}{\pi}\frac{n\Delta s}{1+\tau^{*2}s_k^2}, \quad s_k = k \Delta s \quad \Longrightarrow \phi(\tau) \approx \frac{n}{\tau^*}e^{-|\tau|/\tau^*} \quad \mbox{for }|\tau| \ll \tau_{\rm rec}:=\frac{2\pi}{\Delta s}
                \end{equation}
            \end{subequations}
			with positive real numbers $n$, $\tau^*$, and $\Delta s$. Here, $\tau_{\rm rec}$ is the recurrence time due to the non-zero infrared cutoff $\Delta s\neq 0$. In the small infrared-cutoff limit $\Delta s \to 0$ with $K\to \infty$ and $s_{\max}:=K\Delta s \to \infty$, this finite-dimensional model recovers the original infinite-dimensional model. The characteristic timescale of equilibrium relaxation is roughly estimated by $\tau_{\rm eq}\approx \max\{1/\lambda_0,\tau^*\}$. 

        \subsubsection*{Technical remark 1: the recurrence time}
			Here, a technical point is related to the ``recurrence'' time $\tau_{\rm rec}:=2\pi/\Delta s$ due to the non-zero infrared cutoff. Let us consider the previous finite-dimensional model~\eqref{eq:finiteD-nonMarkovRW-exp}, where the approximated memory function $\phi(\tau)$ has an artificial periodic behaviour for $\tau \gtrsim \tau_{\rm rec}$ due to finite $K$. Thus, one should consider the timescale $\tau_{\rm eq}\ll t\ll \tau_{\rm rec}$ when one wants to numerically observe the equilibrium relaxation process of this model; beyond the recurrence time, the approximately embedded model is no longer consistent with the original non-Markov model. This technical point is essentially present even for Zwanzig's model~\cite{ZwanzigB}---an established microscopic model that leads to the GLE in the thermodynamic limit\footnote{
                For Zwanzig's model, the memory kernel of the GLE is given by 
                \begin{equation}
                    \phi(\tau) = \sum_{k=1}^K M_k \cos(s_k \tau), \quad 
                    \frac{\dd \hv_t}{\dd t} = -\int_0^\infty \dd \tau \phi(\tau)\hv_{t-\tau} + \heta, \quad \la\heta_{t_1}\heta_{t_2}\ra \propto \phi(|t_2-t_1|),
                \end{equation}
                where $K$ is the total number of harmonic oscillators. To understand the relaxation dynamics of the GLE, we must focus on the timescale $\tau_{\rm eq}\ll t \ll \tau_{\rm rec}$, where $\tau_{\rm eq}$ is the characteristic timescale of the memory function $\phi(\tau)$ and $\tau_{\rm rec}$ is the recurrence time due to finite $K$. To make the recurrence time infinite $\tau_{\rm rec}\to \infty$, one must consider the thermodynamic limit $\Delta s\to 0$, jointly with $K\to \infty$ and $K\Delta s\to \infty$. 
            }. In this sense, our framework can be validated by considering an appropriate finite-dimensional model within the recurrence time. 

		\subsubsection*{Technical remark 2: the Ginzburg--Landau memory-bath energy}
			We also discuss the relation between the finite-dimensional setup and the Ginzburg--Landau memory-bath energy~\eqref{eq:gl-energy}. This discussion shows that the Ginzburg--Landau memory-bath energy is obtained as the thermodynamic limit of a discrete description. 
			
			We first assume that the wave numbers are given by $s_k =k\Delta s$ with $\Delta s:= 2\pi/L$ for a finite period $L>0$. The discrete version of the Ginzburg--Landau memory-bath energy is given by 
			\begin{equation}
				E_{\rm nctrl}(\Gamma_t) := \int_{-L/2}^{L/2} \left[\kappa\left(\frac{\pd B_t(l)}{\pd l}\right)^2 + U(B_t(l))\right]\dd l,\quad
				B_t(l):=\sum_{k=1}^KM_k \left[w_{k;t}\cos(s_k l)+ z_{k;t}\sin(s_k l)\right].
			\end{equation}
			Under this construction, the field $B_t(l)$ satisfies the periodic boundary conditions
			\begin{equation}
				B_t(l+L) = B_t(l), \quad \frac{\pd B_t(l+L)}{\pd l} = \frac{\pd B_t(l)}{\pd l}.
			\end{equation}

			We now show that the canonical PDF generated from $E_{\rm nctrl}(\Gamma_t)$ solves the advective part of the finite-dimensional fME~\eqref{eq:finiteD-fME}. Indeed, we have 
			\begin{align}
				\mcL_{\rm A}^{(K)}B(l)
				&= \sum_{k=1}^K M_k \left[s_kz_k\cos(s_k l)-s_kw_k\sin(s_k l)\right]
				= \frac{\pd B(l)}{\pd l},\quad 
				\mcL_{\rm A}^{(K)}:=\sum_{k=1}^K \left[s_k z_k\frac{\pd}{\pd w_k} -s_k w_k\frac{\pd }{\pd z_k}\right].
			\end{align}
			Thus, the advective operator generates translations of the field $B(l)$ along the lag coordinate $l$. Therefore, 
			\begin{align}
				\mcL_{\rm A}^{(K)}E_{\rm nctrl}(\Gamma)
				&= \int_{-L/2}^{L/2}\dd l
				\left[
					2\kappa\frac{\pd B(l)}{\pd l}\frac{\pd^2 B(l)}{\pd l^2}
					+U'(B(l))\frac{\pd B(l)}{\pd l}
				\right] \notag \\
				&= \int_{-L/2}^{L/2}\dd l
				\frac{\pd}{\pd l}
				\left[
					\kappa\left(\frac{\pd B(l)}{\pd l}\right)^2+U(B(l))
				\right] 
				= \left[
					\kappa\left(\frac{\pd B(l)}{\pd l}\right)^2+U(B(l))
				\right]_{-L/2}^{L/2}
				=0,
			\end{align}
			where we used the periodic boundary conditions $B(l+L)=B(l)$ and $\pd B(l+L)/\pd l = \pd B(l)/\pd l$. Since the controllable part $E_{\rm ctrl}(x,\bm{a})$ is independent of $\{z_k,w_k\}_{k=1}^K$, we have $\mcL_{\rm A}^{(K)}E_{\rm tot}(\Gamma,\bm{a})=0$ and $\mcL_{\rm A}^{(K)}P_{{\rm can};\bm{a}}(\Gamma)=0$ for $P_{{\rm can};\bm{a}}(\Gamma)\propto e^{-\beta E_{\rm tot}(\Gamma,\bm{a})}$. The jump part of the fME vanishes under Assumptions 1 and 2, as in Eq.~\eqref{eq:sol-fPDE-ME-jump-disappearance}. Thus, the Ginzburg--Landau memory-bath energy is consistent with the finite-dimensional fME~\eqref{eq:finiteD-fME}.

			By taking the simultaneous thermodynamic and continuum limits $L\to\infty$ and $K\to\infty$, with $\Delta s=2\pi/L\to0$, $s_{\max}:=K\Delta s=2\pi K/L\to\infty$, and $M_k=M(s_k)\Delta s$, we recover the continuous-field description of the Ginzburg--Landau memory-bath energy~\eqref{eq:gl-energy}. 

    \subsection{On the previous formulation by Speck and Seifert~\cite{SpeckSeifert2007}}
        Since a prominent previous formulation of non-Markov stochastic thermodynamics was developed by Speck and Seifert~\cite{SpeckSeifert2007}, we make a short remark about the connection between Ref.~\cite{SpeckSeifert2007} and our work. The starting point of Ref.~\cite{SpeckSeifert2007} is a standard time-local master equation: 
        \begin{equation}
            \frac{\pd P_t(x)}{\pd t}=L(t;t_0)P_t(x),\label{eq:SpeckSeifertME}
        \end{equation}
        where $P_t(x)$ is the single-time-point PDF for the non-Markov dynamics $\hx_t$ and $L(t;t_0)$ is a ``substitute'' operator for non-Markov process~\cite{HanggiZPhys1977}. Speck and Seifert provided a non-Markov framework for stochastic thermodynamics, assuming the existence of the single-time-point master equation~\eqref{eq:SpeckSeifertME}. 

        Here, the existence of the substitute operator $L(t;t_0)$ is a delicate assumption; while \cite{SpeckSeifert2007} explicitly showed the substitute operator for the GLE, it is also known that such an operator is not always well-defined. Indeed, Ref.~\cite{HanggiJSP1978} provided a counterexample where the substitute operator cannot be defined even for the GLE, illustrating that the existence of $L(t;t_0)$ is itself a delicate and nontrivial assumption. Since the conditions under which the substitute operator exists are not known in general, the precise applicability range of Speck--Seifert's formulation is not clear a priori. This contrasts with our framework, whose applicability is stated explicitly through Fourier embeddability and Assumptions 0--3.
     
    \subsection{Open question 1: beyond the Zwanzig jump model}
		In this article, we explicitly construct the ZJM as a thermodynamically consistent Fourier-embeddable class of non-Markov processes that can be captured by stochastic thermodynamics. We then show two concrete examples (a non-Markov two-state system and a non-Markov random-walk model). These models go beyond the GLE and semi-Markov frameworks because their intensities depend on the entire jump history. For even parity $\eps=+1$, their microscopic foundation is clear from the viewpoint of classical Hamiltonian dynamics. Indeed, they can be interpreted as a jump version of the Zwanzig model~\cite{ZwanzigB}.

		On the other hand, Fourier-embeddable non-Markov jump models beyond the ZJM form a conceptually important class for stochastic thermodynamics. Our formulation accommodates a substantially wider class than the ZJM, provided that the additional Assumptions 0--3 are satisfied; the Ginzburg--Landau memory-bath energy~\eqref{eq:gl-energy} is one promising candidate. Exploring this wider class and its microscopic foundation is an important future direction.
		
	\subsection{Open question 2: How large is the Fourier-embeddable class?}
			Another open question is how large a class can be covered by the Fourier-embedding formulation. In our formulation, we have assumed that \vspace{-1mm}
		\begin{itemize}
			\item The history dependence in the intensity functional $\lambda_{\bm{a}}[y\mid \hx_t, \{\hv_{\tau}\}_{t_0\leq \tau \leq t}, \hH_0]$ factors only through the Fourier-auxiliary variables, such that $\lambda_{\bm{a}}[y\mid \hx_t, \{\hv_{\tau}\}_{t_0\leq \tau \leq t}, \hH_0]=\lambda_{\bm{a}}[y\mid \hGamma_t]$. \vspace{-1mm}
			\item The bath initial condition is specified by Eq.~\eqref{eq:init-condition-H0}, which is sampled from the canonical distribution given $\hx_{t_0}$. \vspace{-1mm}
			\item The stationary PDF $P_{\rm ss}[\Gamma]$ exists after the Fourier embedding.\vspace{-1mm}
		\end{itemize}
		These conditions implicitly restrict the model class. Exploring the details of the Fourier-embeddable class and beyond is an interesting future topic.

    \subsection{Open question 3: relationship with stochastic thermodynamics for the semi-Markov processes}\label{sec:conclusion-OQ3}
        Before our work, stochastic thermodynamics for non-Markov processes has not been well explored except for a few specific classes. One important exception is the semi-Markov process~\cite{KlafterB}: a jump process depending only on the previous state but with a general waiting-time distribution. The ``necessary and sufficient'' conditions for the time-reversal symmetry of the semi-Markov process were formulated in Refs.~\cite{Chari1994,Wang2007}, where the detailed-balance condition for the embedded Markov chain and the direction-time independence (DTI) are crucial. A stochastic-thermodynamic framework was also developed in Ref.~\cite{Andrieux2008,Esposito2008,MaesSemiMarkov2009} for the semi-Markov processes based on the generalised detailed-balance (GDB) condition and the DTI assumption. 

        However, there remain controversial points in the community because recent studies show that the GDB and DTI formulation is not a unique framework in developing physically-reasonable stochastic thermodynamics. Indeed, van der Meer et al.~\cite{JannPRX2022} presented a clear counterexample:  \vspace{-1mm}
        \begin{itemize}
            \item They start from a Markov jump process with time-reversal symmetry; thus, conventional Markov stochastic thermodynamics works trivially. \vspace{-2mm}
            \item By applying a coarse-graining, the Markov jump model reduces to a semi-Markov process. \vspace{-2mm}
            \item The reduced semi-Markov process does not satisfy the DTI assumption, thus violating the ``necessary and sufficient'' condition for the time-reversal symmetry. \vspace{-2mm}
            \item This implies that the ``necessary and sufficient'' condition for the semi-Markov time-reversal symmetry is not universally necessary from a physical viewpoint because the semi-Markov model might be a coarse-grained description of another truly microscopic model that is modelled by a conventional Markov process. \vspace{-1mm}
        \end{itemize}
        Remarkably, the entropy production based on the semi-Markov description should be regarded as just a coarse-grained entropy production in the specific case of Ref.~\cite{JannPRX2022}, but not a true physical one. 
        
		This counterexample highlights that defining the time-reversal symmetry of a semi-Markov process is not physically unique; there are several parallel and independent formulations. See Refs.~\cite{Blom2024,Dieball,Martinez2019} for semi-Markov processes employing different notions of time reversal, each of which therefore has its own corresponding identification of entropy production (see Sec. VI D of van der Meer et al.~\cite{JannPRX2022} for a deeper discussion). Thus, there is a fundamental ambiguity about how to define the ``true'' time-reversal symmetry and therefore also how to define entropy production, particularly when the ``truly microscopic'' description of a system is not clear a priori. If the semi-Markov model represents the truly microscopic description, then the framework employing GDB and DTI~\cite{Wang2007} appears the most promising, because it straightforwardly extends the Markovian case. However, because most physical stochastic processes are just coarse-grained descriptions of truly microscopic dynamics (e.g., the Hamiltonian dynamics), the GDB and DTI framework is not always the necessary solution.

		While there is no consensus on how to build a unique stochastic-thermodynamic framework even for semi-Markov processes, our work presents a general framework for Fourier-embeddable non-Markov jump processes. Developing a single unifying framework would be a valuable future direction.

\section{Conclusion}\label{sec:conclusion}
	In this work, we introduce the Fourier embedding framework and derive the corresponding field master equation for the Fourier-embeddable non-Markov jump processes. We study the conditions for thermodynamic consistency (Assumptions 1 and 2) and analyse the mathematical properties of our field master equation. We then develop stochastic thermodynamics for Fourier-embeddable non-Markov jump processes and formulate the first and second laws of stochastic thermodynamics. Under Assumptions 1--3, we show that the thermodynamic laws hold. For equilibrium-to-equilibrium transitions, the cumulative entropy production is determined by the target-system statistics and is therefore gauge invariant under the choice of equilibrium Markov embedding. As concrete examples, we study the Zwanzig jump models (ZJMs), which include a two-state Zwanzig jump model and a random-walk Zwanzig jump model. Remarkably, for even parity $\eps=+1$, the Zwanzig jump models have a microscopic foundation, having a parallel structure to the original Zwanzig model. We numerically verify the second law for these two models. 

\appendix        
\section{Time-reversal symmetry}\label{sec:app:detailedbalance}
	In this Appendix, the necessary and sufficient condition of the time-reversal symmetry is reviewed according to the textbook by Gardiner~\cite{GardinerB}, particularly for the finite-dimensional stochastic process. The time-reversal symmetry for the fME (belonging to an infinite-dimensional case) is formulated as a straightforward generalization of the time-reversal symmetry for the finite-dimensional case. 

	\subsection{Finite-dimensional case}\label{sec:app:detailedbalance:finiteD}
		Let us consider the finite-dimensional Markovian stochastic process described by the PDF $P_t(\bm{x})$ with a finite-dimensional vector argument $\bm{x}:=(x_1,\dots,x_K)$ for a positive integer $K>0$. Generally, the standard differential form of the Chapman--Kolmogorov equation (i.e.\ the ME) is given by 
		\begin{equation}
			\frac{\partial P_t(\bm{x})}{\partial t} = \left[-\sum_{k} \frac{\partial}{\partial x_k}A_k(\bm{x})+\frac{1}{2}\sum_{k,l}\frac{\partial^2}{\partial x_k \partial x_l}B_{kl}(\bm{x})\right]P_t(\bm{x}) + \int \dd \bm{x}'\left[\lambda(\bm{x}\mid \bm{x}')P_t(\bm{x}')-\lambda(\bm{x}'\mid \bm{x})P_t(\bm{x})\right],
		\end{equation}
		with the advective term $A_k$, the positive-semidefinite diffusion matrix $B(\bm{x}):=[B_{kl}(\bm{x})]$, and the jump-intensity density $\lambda(\bm{x}'\mid \bm{x})\geq 0$ from $\bm{x}$ to $\bm{x}'$. 

		The parity variable $\eps_k$ for the $k$th component of $\bm{x}$ is defined by $\eps_k = 1$ if $x_k$ is an even variable, or $\eps_k = -1$ if $x_k$ is an odd variable. The necessary and sufficient condition of the time-reversal symmetry is given by 
		\begin{subequations}
			\label{eq:app:detailed-balance}
		\begin{equation}
			\begin{gathered}
				\lambda(\bm{x}\mid \bm{x}')P_{\rm ss}(\bm{x}') = \lambda(\bm{x}'^* \mid  \bm{x}^*)P_{\rm ss}(\bm{x}), \quad 
				\eps_k\eps_l B_{kl}(\bm{x}^*) = B_{kl}(\bm{x}), \quad P_{\rm ss}(\bm{x})=P_{\rm ss}(\bm{x}^*), \\
				\eps_k A_k (\bm{x}^*) P_{\rm ss}(\bm{x}) = -A_k(\bm{x})P_{\rm ss}(\bm{x}) + \sum_l \frac{\partial}{\partial x_l}[B_{kl}(\bm{x})P_{\rm ss}(\bm{x})] 
			\end{gathered}
		\end{equation}
		for the time-reversal state defined by
		\begin{equation}
			\bm{x}^* :=(\eps_1 x_1, \dots, \eps_K x_K).
		\end{equation}
		\end{subequations}

	\subsection{Infinite-dimensional case}
		Let us consider an fME without diffusive terms:
		\begin{equation}
			\frac{\partial P_t[z]}{\partial t} = -\int_0^\infty \dd s \frac{\delta}{\delta z(s)}A_s[z]P_t[z] + \int \dd z'\left(\lambda[z \mid z']P_t[z']-\lambda[z'\mid z]P_t[z]\right),
		\end{equation}
		where $P_t[z]:=P_t[\{\hz_t(s)=z(s)\}]$ is a probability density functional regarding the auxiliary field variables $\{\hz_t(s)\}_s$. For simplicity, we removed the diffusion term in the fME while keeping the advective and jump terms. By considering the condition~\eqref{eq:app:detailed-balance} for the finite-dimensional ME, the necessary and sufficient condition of the time-reversal symmetry for the fME is then given by 
		\begin{equation}
			\lambda[z\mid z']P_{\rm ss}[z'] = \lambda[z'^* \mid  z^*]P_{\rm ss}[z] , \>\>\>
			\eps_s A_s [z^*] P_{\rm ss}[z] = -A_s[z]P_{\rm ss}[z], \quad P_{\rm ss}[z]=P_{\rm ss}[z^*],
			\label{eq:app:detailed-balance-fME}
		\end{equation}		
		where the time-reversal state is given by 
		\begin{equation}
			z^*(s) = \eps_s z(s).
		\end{equation}		
		When several auxiliary fields are present, as for the pair $\{z(s),w(s)\}_s$ in the fME~\eqref{eq:field_master_hdCP}, the same conditions apply componentwise, with the drifts $A_{z(s)}[\Gamma]$, $A_{w(s)}[\Gamma]$ and the corresponding parities $\eps_z$, $\eps_w$.